\pdfoutput=1

\documentclass[11pt,twoside,a4paper,cmspaper,final,collab]{cms-tdr}
\def\svnVersion{88929:92618}\def\cmsCernNo{2011-217}\def\cmsCernDate{\today}\def\cmsMessage{Submitted to Physics Letters B}

\begin{document}\cmsNoteHeader{TOP-11-014}

\hyphenation{had-ron-i-za-tion}
\hyphenation{cal-or-i-me-ter}
\hyphenation{de-vices}

\RCS$Revision: 92618 $
\RCS$HeadURL: svn+ssh://alverson@svn.cern.ch/reps/tdr2/papers/TOP-11-014/trunk/TOP-11-014.tex $
\RCS$Id: TOP-11-014.tex 92618 2011-12-22 14:23:01Z alverson $
\newcommand{\DeltaEta}{\ensuremath{\Delta\!\left|\eta\right|}} % alternative to definining a delta operator
\cmsNoteHeader{TOP-11-014} % This is over-written in the CMS environment: useful as preprint no. for export versions
\title{Measurement of the charge asymmetry in top-quark pair production in proton-proton collisions at $\sqrt{s} = 7\,\mathrm{TeV}$}

\date{\today}

\abstract{The difference in angular distributions between top quarks and antiquarks, commonly referred to as the charge asymmetry, is measured in pp collisions at the LHC with the CMS experiment. The data sample corresponds to an integrated luminosity of $1.09\,\mathrm{fb}^{-1}$ at a centre-of-mass energy of 7\,TeV. Top-quark pairs are selected in the final state with an electron or muon and four or more jets. At least one jet is identified as originating from b-quark hadronization. The charge asymmetry is measured in two variables, one based on the pseudorapidities ($\eta$) of the top quarks and the other on their rapidities ($y$). The results $A_{C}^{\eta} = -0.017 \pm 0.032~(\mathrm{stat.})\,^{+0.025}_{-0.036}~(\mathrm{syst.})$ and $A_{C}^{y} = -0.013 \pm 0.028~(\mathrm{stat.})\,^{+0.029}_{-0.031}~(\mathrm{syst.})$ are consistent within uncertainties with the standard-model predictions.}

\hypersetup{%
pdfauthor={CMS Collaboration},%
pdftitle={Measurement of the charge asymmetry in top-quark pair production in proton-proton collisions at sqrt(s) = 7 TeV},%
pdfsubject={CMS},%
pdfkeywords={CMS, physics, top quark}}

\maketitle %maketitle comes after all the front information has been supplied

\section{Introduction}

The top quark is the only fundamental fermion with a mass on the order of the scale of electroweak symmetry breaking, and may therefore play a special role in physics beyond the standard model (BSM). In some BSM theories, top-quark pairs can be produced through the exchange of yet unknown heavy particles, in addition to the production through quark-antiquark annihilation and gluon-gluon fusion. Possible candidates include axigluons~\cite{Antunano:2007da,Frampton:2009rk}, \cPZpr\ bosons~\cite{Rosner:1996eb}, and Kaluza-Klein excitations of gluons~\cite{Ferrario:2008wm,Ferrario:2009ee}. Such new particles can appear as resonances in the \ttbar invariant mass spectrum in $s$-channel production of top-quark pairs. If these hypothetical particles are exchanged in the $t$ or $u$ channels, alternative approaches are needed to search for new top-quark production modes~\cite{AguilarSaavedra:2011hz}. One property of \ttbar production that can be sensitive to the presence of such additional contributions is the difference in angular distributions of top quarks and antiquarks, commonly referred to as the charge asymmetry.

In the standard model (SM), a small charge asymmetry in \ttbar production through quark-anti\-quark annihilation appears in QCD calculations at next-to-leading order (NLO)~\cite{Kuhn:1998jr,Kuhn:1998kw}. The interference between the Born diagram and the box diagram, as well as between initial- and final-state radiation, correlates the flight directions of the top quarks and antiquarks to the directions of motion of the initial quarks and antiquarks, respectively. The asymmetric initial state of proton-antiproton collisions leads to an observable forward-backward asymmetry at the Tevatron, where the top quarks are emitted preferentially along the direction of motion of the incoming protons and the top antiquarks along the direction of the antiprotons. This asymmetry is observable in the difference in rapidity ($y$) of top quarks and antiquarks, $y_{\cPqt} - y_{\cPaqt}$. Recent measurements~\cite{:2011kc,PhysRevD.84.112005} by the CDF and D0 Collaborations report asymmetries that are about two standard deviations larger than the value of about 0.08~\cite{Kuhn:1998jr,Kuhn:1998kw,Bernreuther:2010ny,Kuhn:2011ri} predicted in the SM. At high \ttbar invariant mass ($M_{\ttbar} > 450$\GeVcc), the CDF Collaboration finds an even larger asymmetry relative to the SM prediction~\cite{:2011kc}, while the D0 Collaboration does not observe a significant mass dependence of the asymmetry. These results have led to speculations that the large asymmetry might be generated by additional axial couplings of the gluon~\cite{Gabrielli:2011jf} or by heavy particles with unequal vector and axial-vector couplings to top quarks and antiquarks~\cite{Bauer:2010iq,Cao:2009uz,Shu:2009xf,Arhrib:2009hu,Dorsner:2009mq,Djouadi:2009nb,Djouadi:2011aj, Rodrigo:2010gm,Martynov:2010ed,AguilarSaavedra:2011vw,Jung:2009pi,Cao:2010zb,Bai:2011ed,Ligeti:2011vt}.

Owing to the symmetric initial state of proton-proton collisions at the Large Hadron Collider (LHC), the charge asymmetry does not manifest itself as a forward-backward asymmetry; the rapidity distributions of top quarks and antiquarks are symmetrical around $y = 0$. However, since the quarks in the initial state are mainly valence quarks, while the antiquarks are always sea quarks, the larger average momentum fraction of quarks leads to an excess of top quarks produced in the forward directions. The rapidity distribution of top quarks in the SM is therefore broader than that of the more centrally produced top antiquarks. The same effect is visible in the purely geometrically defined pseudorapidity $\eta = -\ln(\tan\theta/2)$, where $\theta$ is the polar angle relative to the counterclockwise beam axis. The charge asymmetry can be observed through the difference in the absolute values of the pseudorapidities of top quarks and antiquarks, $\DeltaEta = \left|\eta_{\cPqt}\right| - \left|\eta_{\cPaqt}\right|$~\cite{Diener:2009ee}. Another approach is motivated in Ref.~\cite{Jung:2011zv}, where the observable used by the Tevatron experiments ($y_{\cPqt} - y_{\cPaqt}$) is multiplied by a factor that accounts for the boost of the ${\ttbar}$ system, yielding $\Delta y^{2} =  (y_{\cPqt} - y_{\cPaqt})\cdot(y_{\cPqt} + y_{\cPaqt}) = (y_{\cPqt}^{2} - y_{\cPaqt}^{2})$. Using either of the two variables, the charge asymmetry can be defined as

\begin{equation}
A_{C} = \frac{N^{+} - N^{-}}{N^{+} + N^{-}}~,
\label{eq:ACDC}
\end{equation}

where $N^{+}$ and $N^{-}$ represent the number of events with positive and negative values in the sensitive variable, respectively. Since the SM charge asymmetry is a higher-order effect in quark-antiquark annihilation, and at the LHC the top-quark pairs are produced mainly through gluon-gluon fusion, the asymmetry expected in the SM at the LHC is smaller than at the Tevatron. For a centre-of-mass energy of 7\TeV, the NLO prediction for an asymmetry in the $\DeltaEta$ variable is $A_{C}^{\eta}(\text{theory}) = 0.0136\pm 0.0008$~\cite{Kuhn:2011ri}, while $A_{C}^{y}(\text{theory}) = 0.0115\pm 0.0006$~\cite{Kuhn:2011ri} for the rapidity variable. The existence of new sources of physics with different vector and axial-vector couplings to top quarks and antiquarks could enhance these asymmetries up to a maximum of 0.08~\cite{AguilarSaavedra:2011hz}. The uncertainties on the above predictions reflect the variations from the choice of  parton distribution functions and different choices of factorization and renormalization scales, as well as the dependence on the top-quark mass within its experimental uncertainty.

For the sake of simplicity, we focus on $\DeltaEta$ when describing the method but quote uncertainties and results for both variables. We discuss below the experimental setup (Section~\ref{sec:cmsDec}), the
data sample (Section~\ref{sec:DataSim}), and the selection of \ttbar candidate
events (Section~\ref{sec:EvtSel}). This is followed by a description of the estimation of the background
contamination (Section~\ref{sec:BkgEst}). The reconstruction of the
kinematics of the top-quark candidates is described in Section~\ref{sec:Reco}. The details of
the applied unfolding and measurement procedures and of the different sources of
systematic uncertainties are given in Section~\ref{sec:Measurement} and Section~\ref{sec:Systematics}, respectively, and the results of the analysis are presented in Section~\ref{sec:Results}.

\section{The CMS detector}
\label{sec:cmsDec}

The central feature of the Compact Muon Solenoid (CMS) apparatus is a superconducting solenoid of 6~m internal diameter, providing a field of 3.8\unit{T}. Within the field volume are the silicon pixel and strip tracker, the crystal electromagnetic calorimeter (ECAL), and the brass/scin\-til\-la\-tor hadron calorimeter (HCAL).
The inner tracker measures trajectories of charged particles within the pseudorapidity range $\left|\eta\right| < 2.5$. It consists of 1440 silicon pixel and 15\,148 silicon strip detector modules and provides an impact parameter resolution of ${\sim}15\mum$ and a transverse momentum ($\pt$) resolution of about 1.5\% for 100\GeVc particles.

The ECAL consists of nearly 76\,000 lead tungstate crystals that
provide coverage in pseudorapidity of $\left|\eta \right|< 1.48$ for the
ECAL barrel region and~$1.48 <\left| \eta \right| < 3.0$ for the two endcaps. A
preshower detector consisting of two planes of silicon
sensors interleaved with a total of three radiation lengths of lead is located in
front of the endcaps. The ECAL energy resolution is 3\% or better for
the range of
electron energies relevant for this analysis. The HCAL is composed of layers of plastic
scintillator within a brass/stainless steel absorber, covering the
region $\left|\eta\right|<3.0$. In the region $\left| \eta \right\vert< 1.74$, the HCAL cells have widths of 0.087 in pseudorapidity and 0.087\unit{rad} in azimuth ($\phi$). In
the $(\eta,\phi)$ plane, for $\left| \eta \right|< 1.48$, the HCAL cells
match the corresponding $5 \times 5$ ECAL crystal arrays to form
calorimeter towers
projecting radially outwards from the centre of the detector. At larger values of $\left| \eta \right|$, the coverage in $\eta$ of each tower increases, although the matching ECAL arrays contain fewer crystals.

Muons are measured in the pseudorapidity range $\left|\eta\right|< 2.4$, with detection planes made using three technologies, drift tubes, cathode strip chambers, and resistive plate chambers, all embedded in the steel return yoke. Matching the muons to the tracks measured in the silicon tracker provides a transverse momentum resolution between 1 and 5\%, for $\pt$ values up to 1\TeVc.
In addition to barrel and endcap detectors, CMS has extensive forward calorimetry. A detailed description of CMS can be found in Ref.~\cite{cms}.
\section{Data and simulation}
\label{sec:DataSim}

This analysis of $\ttbar$ events produced in proton-proton collisions at a centre-of-mass energy of 7\TeV is based on data taken with the CMS detector, corresponding to an integrated luminosity of $1.09\pm 0.04\fbinv$. To translate the distributions measured with reconstructed objects to distributions for the underlying quarks, we use simulated data samples. Top-quark pair events are generated with the tree-level matrix-element generator \MADGRAPH version~5~\cite{madgraph},
interfaced to \PYTHIA version 6.4~\cite{pythia} for the parton showering, where the MLM algorithm~\cite{MLM} is used for the matching.
Spin correlation in decays of top quarks is taken into account and higher-order gluon and quark production is described through matrix elements for up to three extra jets accompanying the $\ttbar$ system. Although the higher-order processes leading to the $\ttbar$ charge asymmetry are not taken fully into account in this leading-order (LO) simulation, its usage is still justified by the fact that these processes affect only the production of top quarks and not their decay, and that the simulated events are used only to reconstruct top-quark momenta from their decay products and to correct for resolution effects. We simulate the main SM backgrounds to top-quark pair production using the same combination of \MADGRAPH and \PYTHIA programs. The radiation of up to four jets in weak vector-boson production is simulated through matrix-element-based calculations (these processes are denoted as $\PW{+}\text{jets}$ and $\cPZ{+}\text{jets}$ in the following). The background from electroweak production of single top quarks is also simulated using the \MADGRAPH generator. Multijet background events are generated using \PYTHIA version~6.4. On average, six additional proton-proton interactions (pile-up) per event are observed in the analysed data, and this pile-up contribution is overlaid on the simulated events, all of which are processed through the CMS detector simulation and reconstructed using standard CMS software.

\section{Event selection}
\label{sec:EvtSel}

In the SM, a top quark decays almost exclusively into a \cPqb\ quark and a \PW\ boson. In this measurement we focus on $\ttbar$ events, where one of the \PW\ bosons from the decay of a top-quark pair subsequently decays into a muon or electron and the corresponding neutrino, and the other \PW\ boson decays into a pair of jets. We therefore select events containing one electron or muon and four or more jets, at least one of which is identified as originating from the hadronization of a \cPqb\ quark. For the reconstruction of electrons, muons, jets, and any imbalance in transverse momentum due to the neutrino, we use a particle-flow (PF) algorithm~\cite{PFT-09-001}. This algorithm aims to reconstruct the entire event by combining information from all subdetectors, including tracks of charged particles in the tracker and the muon system, and energy depositions in the electromagnetic and hadronic calorimeters.

We select events with triggers that require an electron or muon with transverse momentum greater than 25\,\GeVc and 17\,\GeVc, respectively, together with at least three jets, each with $\pt > 30\GeVc$. In addition, we require the primary vertex reconstructed from the tracks with the largest summed transverse momentum to be located in a cylindrical region defined by the longitudinal distance $\left| z \right| < 24\unit{cm}$ and radial distance $r < 2\unit{cm}$ relative to the centre of the CMS detector.

In the $\text{electron}{+}\text{jets}$ selection, the electron candidates are required to have a transverse momentum of $\pt > 30\,$\GeVc and to be within the region $\left|\eta\right| <  2.5$, excluding the transition region between the ECAL barrel and endcaps of $1.4442 < |\eta_{\mathrm{sc}}| < 1.5660$, where $\eta_{\text{sc}}$ is the pseudorapidity of the electron candidate's supercluster, which corresponds to the cluster of ECAL energy depositions from the electron and any accompanying bremsstrahlung photons~\cite{ElectronPAS}. The transverse impact parameter of the electron track relative to the beam axis is required to be smaller than $0.02\unit{cm}$. The energy in the HCAL cell that is mapped onto the supercluster must be less than 2.5\% of the total ECAL energy associated with the supercluster. Additional requirements are made on the spatial distribution of the shower and the angular separation between the ECAL supercluster and the matching track. The longitudinal position of the electron track at its closest approach to the beam line is required to lie within $1\unit{cm}$ of the longitudinal position of the primary vertex, to ensure that the electron is emitted from the primary interaction. Also, electron candidates must be isolated. The lepton isolation variable $I_{\text{Rel}}^{\ell}$ is based on the reconstructed energies of particle-flow objects relative to the lepton transverse momentum ($\pt^{\ell}$):

\begin{equation}
I_{\text{Rel}}^{\ell} = \frac{E_{\text{CH}}^{\ell} + E_{\text{NH}}^{\ell} + E_{\gamma}^{\ell}}{\pt^{\ell}\cdot c}~,
\end{equation}

where $E_{\text{CH}}^{\ell}$ is the energy deposited by charged hadrons in a cone with radius $\Delta R = 0.4$ in $(\eta,\phi)$ around the lepton track, and $E_{\text{NH}}^{\ell}$ and $E_{\gamma}^{\ell}$ are the respective energies of neutral hadrons and photons. We require electron candidates to have $I_{\text{Rel}}^{\Pe} < 0.125$. Events with exactly one electron candidate satisfying these quality criteria are selected for further consideration. Electron candidates that lack signals in the inner layers of the tracking system or that can be paired with a second track of opposite curvature are assumed to be the product of photon conversions and are discarded.

Muons are reconstructed using the combined information from the silicon tracker and muon system. In the selection of $\text{muon}{+}\text{jets}$ events, the muons are required to have $\pt > 20\GeVc$ and lie within the muon trigger acceptance ($\left|\eta \right|<2.1$). The same requirements as for electron candidates are imposed on the transverse impact parameter and the longitudinal origin of the muon track. The muon candidate is required to have a prescribed minimum number of hits in both the silicon tracking system and the muon chambers, and must be isolated from other energy depositions in the event, again defined by $I_{\text{Rel}}^{\Pgm} < 0.125$. When more than one muon passes all these criteria, the event is rejected.

Dilepton events from $\ttbar$ and \cPZ-boson decays are suppressed by applying a veto on additional, less stringently defined charged leptons in the event. We reject all events containing any additional electron candidates with $\pt>15\GeVc$, $\left|\eta\right| < 2.5$, and $I_{\text{Rel}}^\Pe<0.25$, or additional muon candidates with $\pt>10\GeVc$, $\left|\eta\right|<2.5$, and $I_{\text{Rel}}^{\Pgm}<0.25$.

We cluster all particles reconstructed through the particle-flow algorithm, excluding isolated electron and muon candidates, into jets using the anti-$k_{\mathrm{T}}$ jet algorithm~\cite{antikt} with the distance parameter $R = 0.5$, as constructed with \textsc{FastJet} version 2.4~\cite{fastjet,fastjet2}.
The jet energy is corrected for additional contributions from multiple interactions, as well as for $\eta$ and $\pt$-dependent detector response. To account for observed differences of about 10\% in jet-energy resolution between data and simulation, a correction is applied to jets in the simulated samples so that their resolutions match those measured in data. Selected jets are required to be within $\left|\eta\right|<2.4$ and have corrected $\pt > 30\GeVc$. At least four jets must be present in an event, and at least one of the jets must be tagged as coming from the hadronization of a b quark by an algorithm that orders the tracks in impact parameter significance and discriminates using the track with the second highest significance~\cite{TCHE,BTAG}. This algorithm has a tagging efficiency of about 60\%, evaluated using \cPqb\ jets containing muons from semileptonic decays of \cPqb\ hadrons~\cite{BTAG}, and a misidentification rate of about 1\%~\cite{BTAG}.

\section{Estimation of background}
\label{sec:BkgEst}

Applying the selection criteria described above, we find a total of 12\,757 events, 5665 in the $\text{electron}{+}\text{jets}$ channel and 7092 in the $\text{muon}{+}\text{jets}$ channel. From an evaluation of the simulated background processes, we expect a background contribution of about 20\% to the selected data sample. We estimate the contributions from the various background processes separately for the two lepton flavours. We make use of the discriminating power of the imbalance in transverse momentum in an event, $\ETmiss$, and of M3, the invariant mass of the combination of three jets that corresponds to the largest vectorially summed transverse momentum~\cite{Chatrchyan:2011ew}. For the $\PW{+}\text{jets}$, $\cPZ{+}\text{jets}$, and single-top-quark background processes, the respective simulated samples are used to model the shapes of the $\ETmiss$ and M3 distributions, while an approach based on data is pursued for the multijet background.

Background contributions are estimated in both channels separately by means of binned max\-imum-likelihood fits to the two distributions. The $\ETmiss$ distribution shows the largest discrimination power for small values, where it discriminates between events with and without neutrinos in the final state. We therefore separate the data sample into events with $\ETmiss < 40\GeV$ and $\ETmiss > 40\GeV$, and simultaneously fit the $\ETmiss$ distribution for the low-$\ETmiss$ sample and the M3 distribution from the high-$\ETmiss$ sample, to obtain estimates of the numbers of events from each process in the entire data sample ($\ETmiss \geq 0$).

The $\ttbar$ signal and all the above-listed background processes enter the likelihood function with a single fit parameter for the normalization of their respective $\ETmiss$ and M3 distributions. $\PW{+}\text{jets}$ production is inherently asymmetric at the LHC, with more $\PWp$ bosons being produced than $\PWm$ bosons. As the distributions of kinematic variables for the two processes are slightly different, this could introduce an artificial contribution to the measured $\ttbar$ charge asymmetry. Therefore, this background process requires special care, and we measure the $\PW{+}\text{jets}$ contributions from $\PWp$ and $\PWm$ bosons separately, using different fit parameters for the two sources of $\PW{+}\text{jets}$.

As mentioned above, for all background processes, with the exception of multijet events, we rely on simulations to model the $\ETmiss$ and M3 distributions. Since the overall cross section for multijet production is several orders of magnitude larger than that of any other process, this specific background can be modelled directly from data by defining an appropriate region enriched in multijet events. In both lepton channels, the largest suppression of multijet events in the default event selection is achieved by requiring isolation of the charged leptons. Consequently, to enrich background from multijet events, we require $0.3 < I_{\text{Rel}}^{\ell} < 0.5$, instead of $I_{\text{Rel}}^{\ell} < 0.125$. To avoid double counting of energy contributions, the momenta of these electron and muon candidates are removed from that of the jet to which they were assigned. The event samples obtained with these altered selections are estimated using simulated events to have multijet purities of 92\% in the $\text{muon}{+}\text{jets}$ channel and 87\% in the $\text{electron}{+}\text{jets}$ channel.

The $\cPZ{+}\text{jets}$ contribution to the selected data is expected to be small, and is difficult to discriminate from the multijet background processes, especially in the $\ETmiss$ distributions. It is also very difficult to discriminate single-top-quark production from the $\ttbar$ signal. Both single-top-quark and $\cPZ{+}\text{jets}$ production are well understood theoretically and their expected contributions are modest. We therefore constrain the numbers of $\cPZ{+}\text{jets}$ and single-top-quark events in the fit to the predictions from simulation, assigning an uncertainty of 30\%, as was done in Ref.~\cite{Chatrchyan:2011ew}, through Gaussian functions in the likelihood. The numbers of events for all other processes are left free in the fit.

\begin{figure}[h!]
 \centering
   \includegraphics[width=0.49\textwidth]{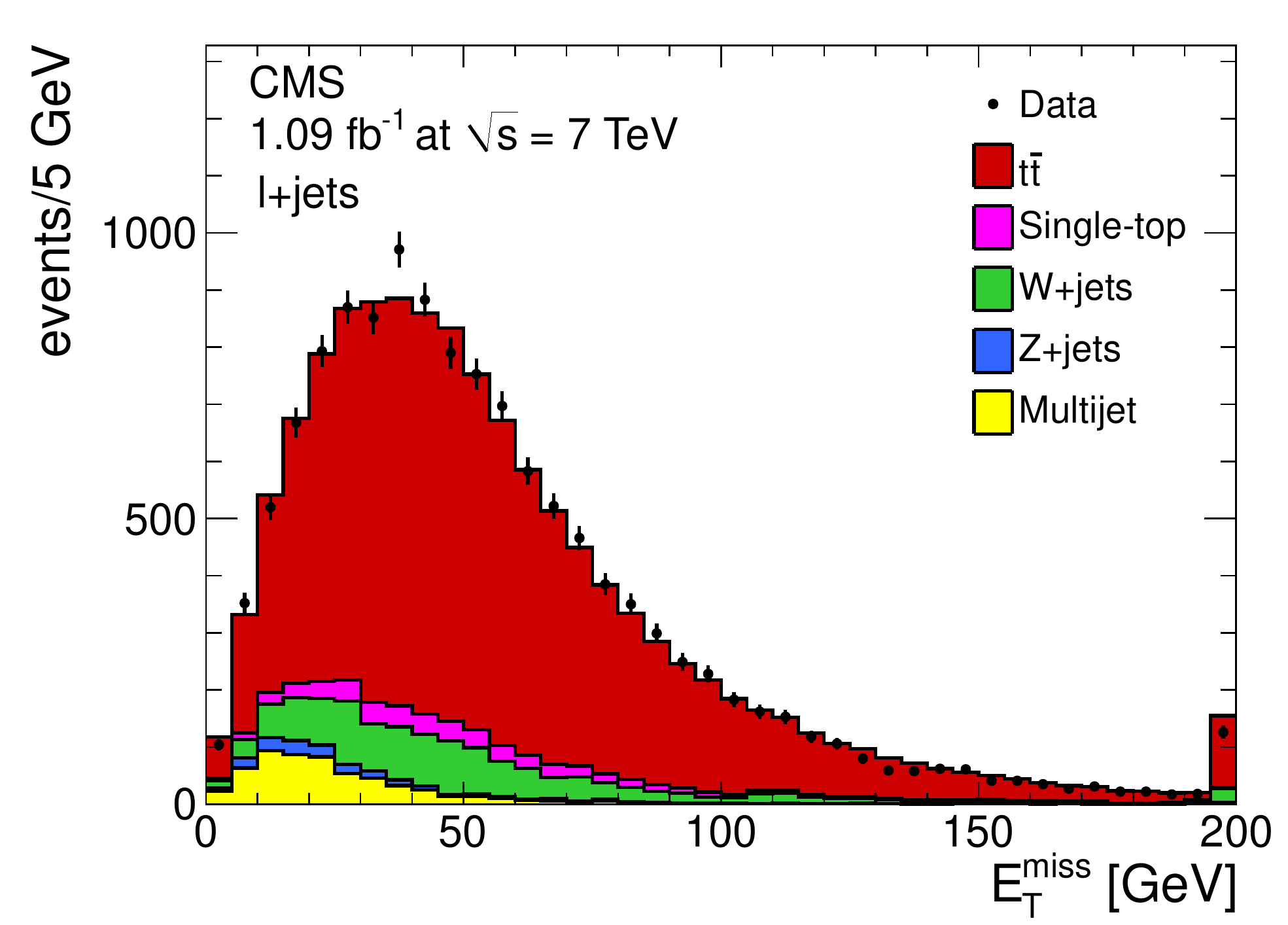}
   \includegraphics[width=0.49\textwidth]{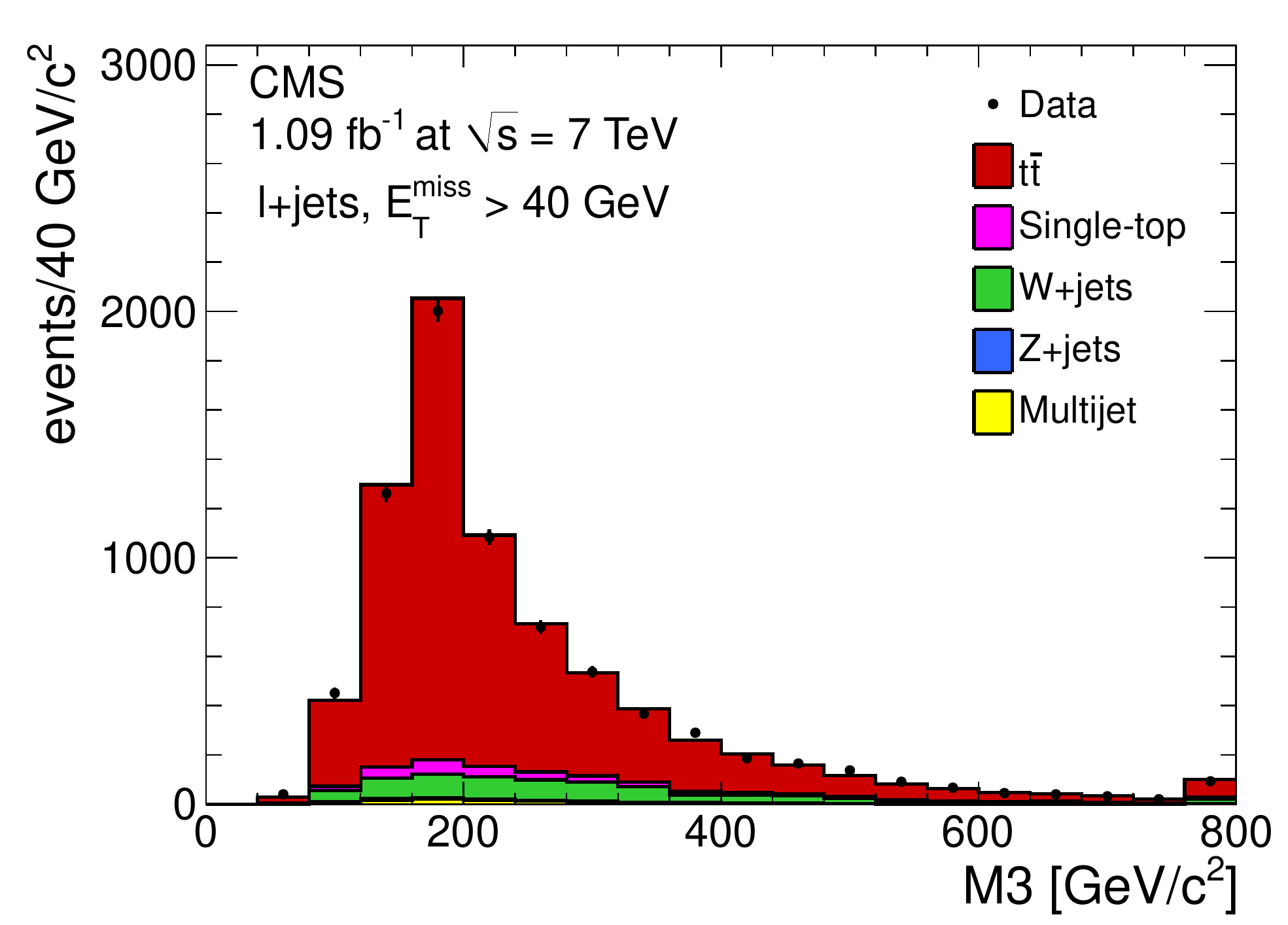}
     \caption{Comparison of the combined $\text{lepton}{+}\text{jets}$ data with simulated contributions for the distributions in $\ETmiss$ (left) and M3 (right). The last bins include the sum of all contributions for $\ETmiss > 200\GeV$ and $\mathrm{M3} > 800\GeVcc$, respectively. The simulated signal and background contributions are normalized to the results of the fits in Table~\ref{tab:FitResults}.}
 \label{fig:DataMCfitresult}
\end{figure}

Table~\ref{tab:FitResults} summarizes the results of the fits separately for the $\text{electron}{+}\text{jets}$ and $\text{muon}{+}\text{jets}$ channels, along with their statistical uncertainties and the sum. The largest correlation is found between the rates for the $\PWp{+}\text{jets}$ and $\PWm{+}\text{jets}$ backgrounds ($+$17\%). The rates of $\PWm{+}\text{jets}$ and $\cPZ{+}\text{jets}$ backgrounds are anticorrelated with that from multijet background ($-$10\%). All other correlations among the fit parameters are found to be small. Figure~\ref{fig:DataMCfitresult} shows the measured $\ETmiss$ and M3 distributions summed for the two channels, with the individual simulated contributions normalized to the results from the fit.

\begin{table*}[htbp]
 \caption{\label{tab:FitResults}Results for the numbers of events for background and $\ttbar$ contributions from fits to data in the $\text{electron}{+}\text{jets}$ and $\text{muon}{+}\text{jets}$ channels, along with their statistical uncertainties. The uncertainties quoted for the single-top-quark and $\cPZ{+}\text{jets}$ backgrounds are related to the constraints used as input for the likelihood fit, and are not the statistical uncertainties from the fit. The last column gives the sum of both channels, where the uncertainties have been added in quadrature. The number of events observed in each channel can be found in the last row.}
 \begin{center}
    \begin{tabular}{lrrr} \hline
      process                          & $\text{electron}{+}\text{jets}$ & $\text{muon}{+}\text{jets}$ &  total~~~~~~~\\    \hline
      Single-top ($\cPqt + \cPqt\PW$)            &$213 \pm 58$~~     &$293 \pm 81$~~   & $506 \pm 99$~~  \\
      $\PWp{+}\text{jets}$                &$313 \pm 84$~~     &$404 \pm 106$    & $717 \pm 135$\\
      $\PWm{+}\text{jets}$                &$299 \pm 90$~~     &$245 \pm 109$    & $544 \pm 141$\\
      $\cPZ{+}\text{jets}$                         &$81 \pm 24$~~      &$85 \pm 26$~~    & $166 \pm 35$~~~\\
      Multijet                         &$355 \pm 71$~~     &$232 \pm 79$~~   & $587 \pm 106$\\ \hline
      Total background                 &$1261 \pm 155 $    &$1259 \pm 191$   & $2520 \pm 246 $      \\
      $\ttbar$                         &$4401 \pm 165$     &$5835 \pm 199$   & $10\,236 \pm 258$\\ \hline
      Observed                         & 5665~~~~~         & 7092~~~~~       & 12\,757~~~ \\ \hline
 \end{tabular}
\end{center}
\end{table*}

\section{Reconstruction of \texorpdfstring{\boldmath{$\mathrm{t\overline{t}}$}}{t-tbar} pairs}
\label{sec:Reco}

The measurement of the \ttbar charge asymmetry is based on the full reconstruction of the four-momenta of the top quark and antiquark in each event. This is done in two steps: first, by reconstructing the leptonically decaying \PW\ boson, and then by associating the measured jets in the event with quarks in the \ttbar decay chain.

The transverse momentum of the neutrino is taken to be the reconstructed \ETm vector. To calculate the longitudinal component of the neutrino momentum, a quadratic constraint, relying on the known mass and decay kinematics of the \PW\ boson, is used. This procedure leads to two solutions for the longitudinal momentum of the neutrino. If these solutions are complex, the transverse components of the neutrino momentum are adjusted such that the \pt of the neutrino is as close as possible to the measured \ETm and the imaginary part of the $p_{z}$ solution vanishes. Adding the resulting four-momentum of the neutrino to that of the charged lepton defines the four-momentum of the parent \PW\ boson. Combining the four-vector of one of the jets in the event with that of the \PW\ boson results in the four-vector of the top quark decaying to the charged lepton in the final state, while the other top quark is reconstructed by combining three of the remaining jets. The charge of the lepton then defines which of the two reconstructed four-momenta corresponds to the top quark, and which to the top antiquark.

From the list of possible reconstructions in each event, we choose the hypothesis that best matches the assumption of a \ttbar interpretation. In simulated \ttbar events, the best possible hypothesis is defined through comparing the reconstructed and true momenta of the top quarks and \PW\ bosons. This kind of information is not accessible in data, and we therefore calculate the probability $\psi$ for each hypothesis to be the best possible one. The calculation of $\psi$ uses the masses of the two reconstructed top quarks and of the hadronically decaying \PW\ boson, as well as the \cPqb-tag  information for the four jets assigned to the four final-state quarks. The three masses are correlated, especially those of the hadronically decaying \PW\ boson and top quark. Assuming a linear correlation, which is confirmed from simulation, they are redefined in terms of three uncorrelated masses $m_1$, $m_2$, and $m_3$, through a rotation matrix derived from simulated \ttbar events. The mass $m_1$ is almost identical to the mass of the top quark decaying to the charged lepton in the final state, while $m_2$ and $m_3$ are mixtures of the masses of the other top quark and the hadronically decaying \PW\ boson. For each of the three uncorrelated masses $m_i$ we calculate a likelihood ratio function $L_{i}(m_i)$, that provides a measure of the probability for a given hypothesis with a certain value of $m_i$ to be the best possible one.

In addition, we consider the \cPqb-tag  values for the jets assigned to the two \cPqb\ quarks and the two light quarks. The probability that a jet with \cPqb-tag  value $x$ is assigned to one of the \cPqb\ quarks in the best possible hypothesis is estimated in simulated \ttbar events and denoted as $P_{b}(x)$. The probability that an assignment of a jet to one of the light quarks is the best possible assignment is then given by $(1- P_b(x))$.

Finally, we choose in each event the hypothesis with the largest value of $\psi$:

\ifthenelse{\boolean{cms@external}}
{
\begin{multline}
\psi = L_{1}(m_1)L_{2}(m_2)L_{3}(m_3)\\
\times P_b(x_{b1})P_b(x_{b2})(1- P_b(x_{q1}))(1 - P_b(x_{q2}))~,
\end{multline}
}
{
\begin{equation}
\psi = L_{1}(m_1)L_{2}(m_2)L_{3}(m_3)P_b(x_{b1})P_b(x_{b2})(1- P_b(x_{q1}))(1 - P_b(x_{q2}))~,
\end{equation}
}
where $x_{b1}$, $x_{b2}$, $x_{q1}$, and $x_{q2}$ are the \cPqb-tag  values for the jets assigned to the two \cPqb\ quarks and two light quarks, respectively.

Studies using simulated \ttbar events show that in about 29\% of all events, we choose the best possible hypothesis using the $\psi$ criterion. In about 72\% of all events, the values of $\DeltaEta$ and $\Delta y^{2}$ are reconstructed with the correct sign.

To check that the simulated background adequately describes the data, several kinematic distributions in data samples without \cPqb-tagged jets, where the dominant contribution is from $\PW{+}\text{jets}$ processes, are compared with those in simulated $\PW{+}\text{jets}$ events. The observed agreement between the measured and the simulated distributions substantiates that the simulated samples used in the analysis describe well the reconstructed quantities in data.
\section{Measurement of the \texorpdfstring{\ttbar}{t-tbar} charge asymmetry}
\label{sec:Measurement}

\begin{figure*}[htbp]
 \centering
   \includegraphics[width=0.49\textwidth]{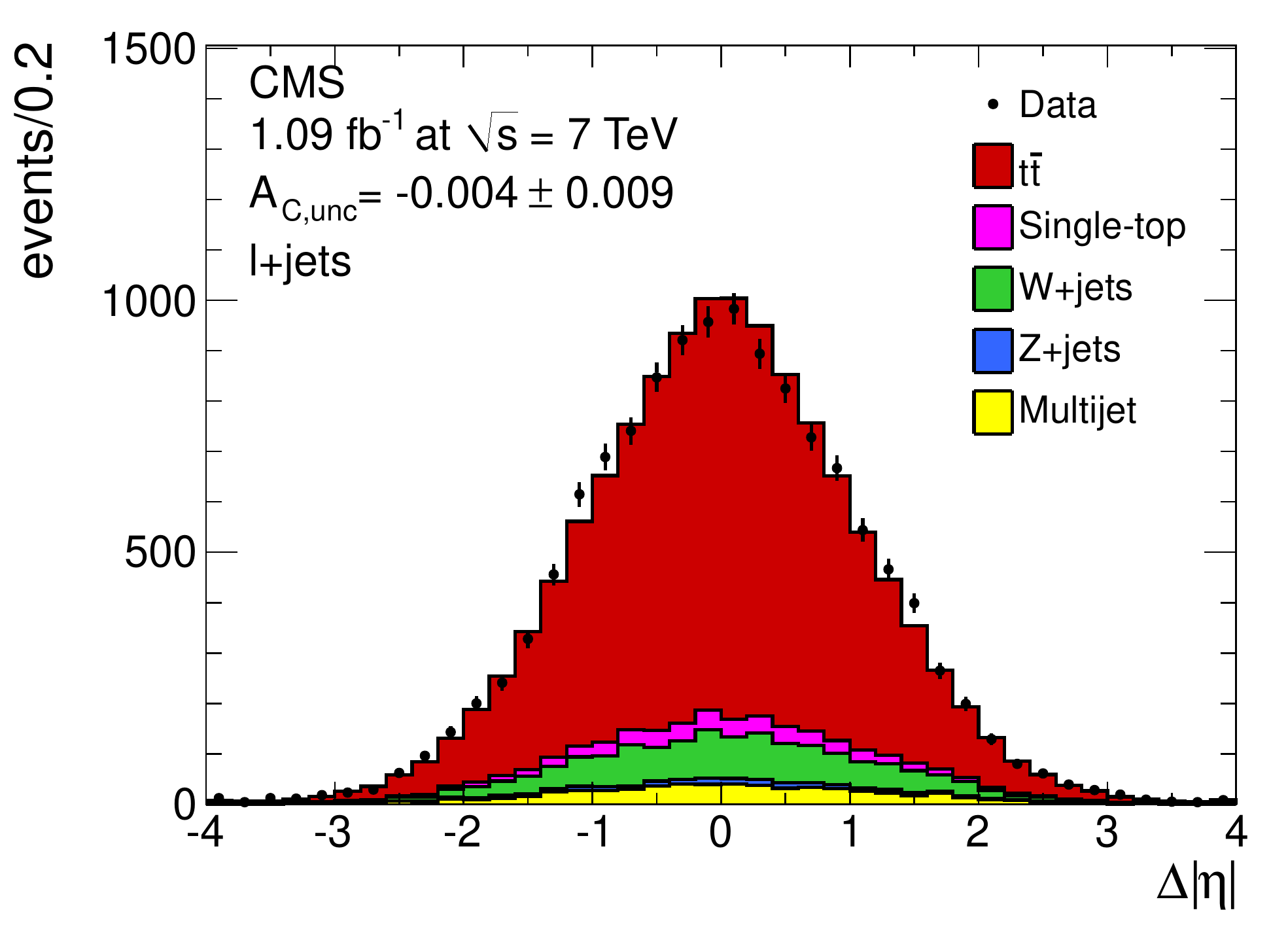}
   \includegraphics[width=0.49\textwidth]{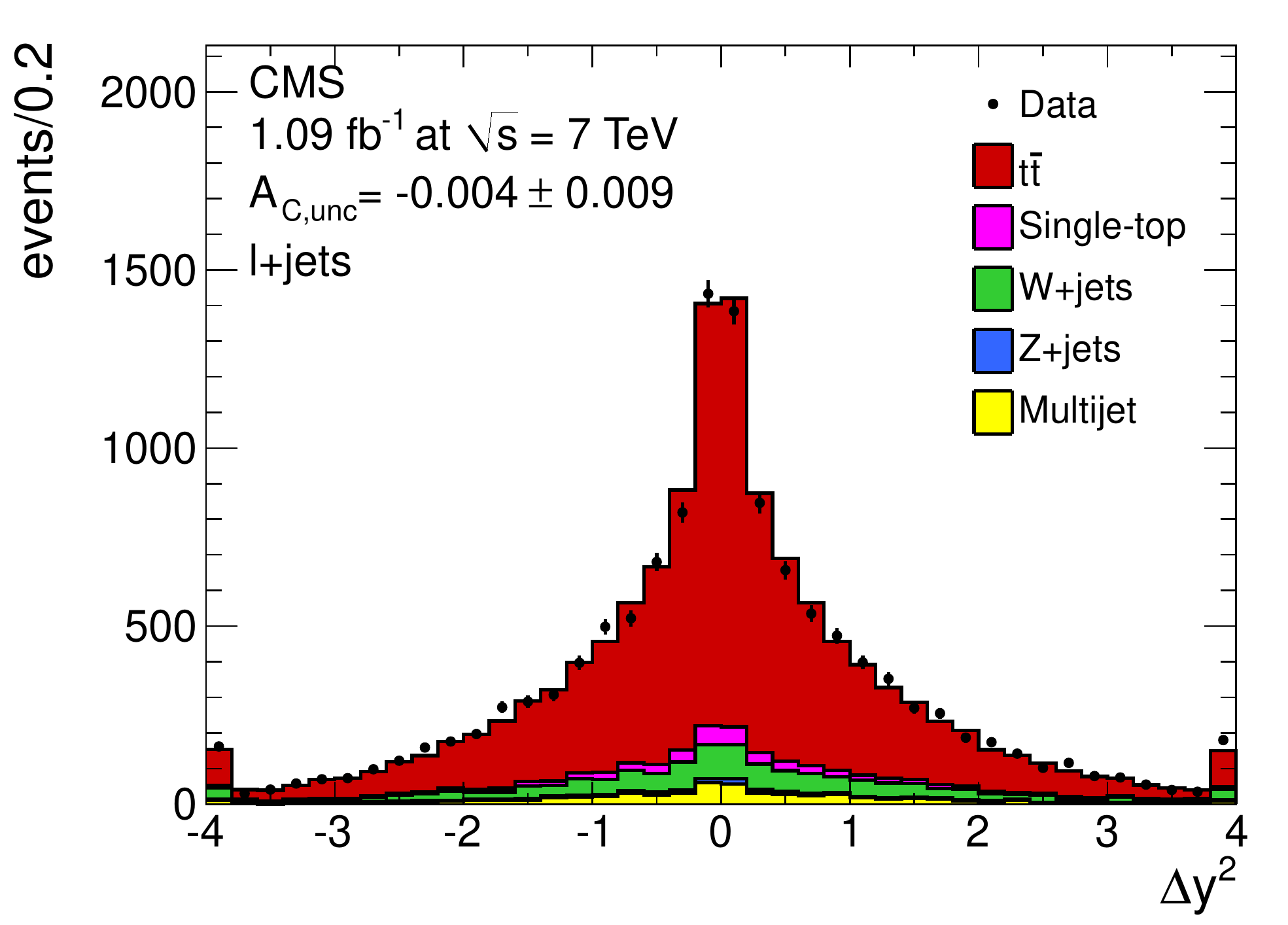}
         \caption{Reconstructed $\DeltaEta$ (left) and $\Delta y^{2}$ (right) distributions for the combined $\text{lepton}{+}\text{jets}$ channel. The last bins include the sum of all contributions for $\left|\DeltaEta\right| > 4.0$ and $\left|\Delta y^{2}\right| > 4.0$, respectively. The signal and background contributions are normalized to the results in Table~\ref{tab:FitResults}.}
 \label{fig:Asy_DataMC}
\end{figure*}

The distributions in the two sensitive variables $\DeltaEta$ and $\Delta y^{2}$ obtained from the reconstructed top quark and antiquark four-vectors are shown in Fig.~\ref{fig:Asy_DataMC}. These distributions are used to calculate an uncorrected charge asymmetry $A_{C,\mathrm{unc}}$ by simply counting the numbers of events with positive and negative values. Using the definition in Eq.~(\ref{eq:ACDC}), we find $A_{C,\mathrm{unc}}^{\eta} = -0.004 \pm 0.009$ and $A_{C,\mathrm{unc}}^{y} = -0.004 \pm 0.009$, where the uncertainties are statistical only.

The above values cannot be compared directly with the theoretical predictions, since several effects bias the measurement at this stage. First, despite the application of relatively stringent \ttbar event selections, about 20\% of all events arise from background processes. The simulated distributions for these background processes exhibit no significant asymmetries. We normalize these distributions to the observed background rates, and subtract them from the data, assuming Gaussian uncertainties on the background rates as well as on statistical fluctuations in the background templates. The effect of the correlations among the individual background rates, discussed in Section~\ref{sec:BkgEst}, is found to be negligible.

\begin{figure*}[htbp]
 \centering
   \includegraphics[width=0.49\textwidth]{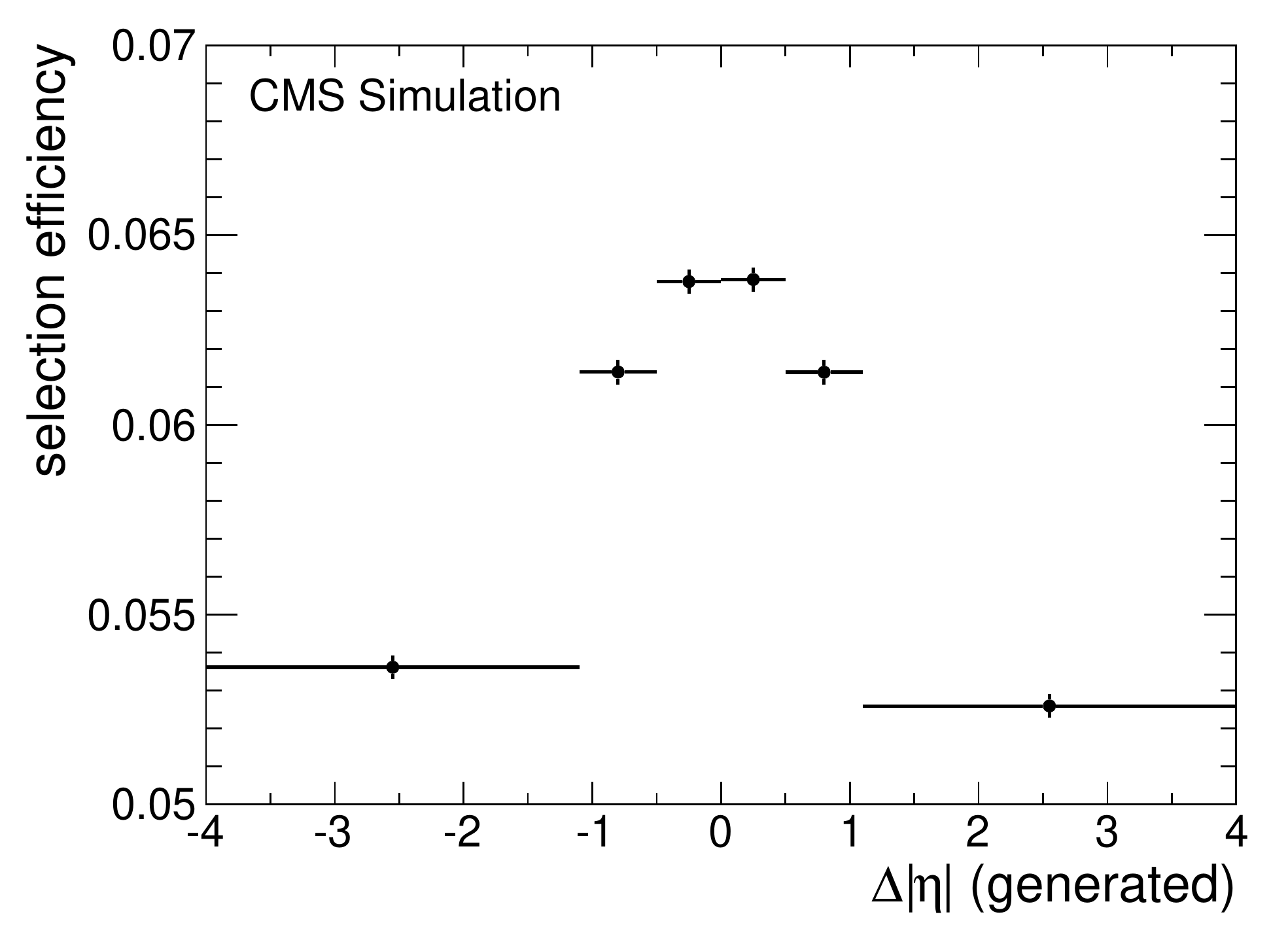}
 \includegraphics[width=0.49\textwidth]{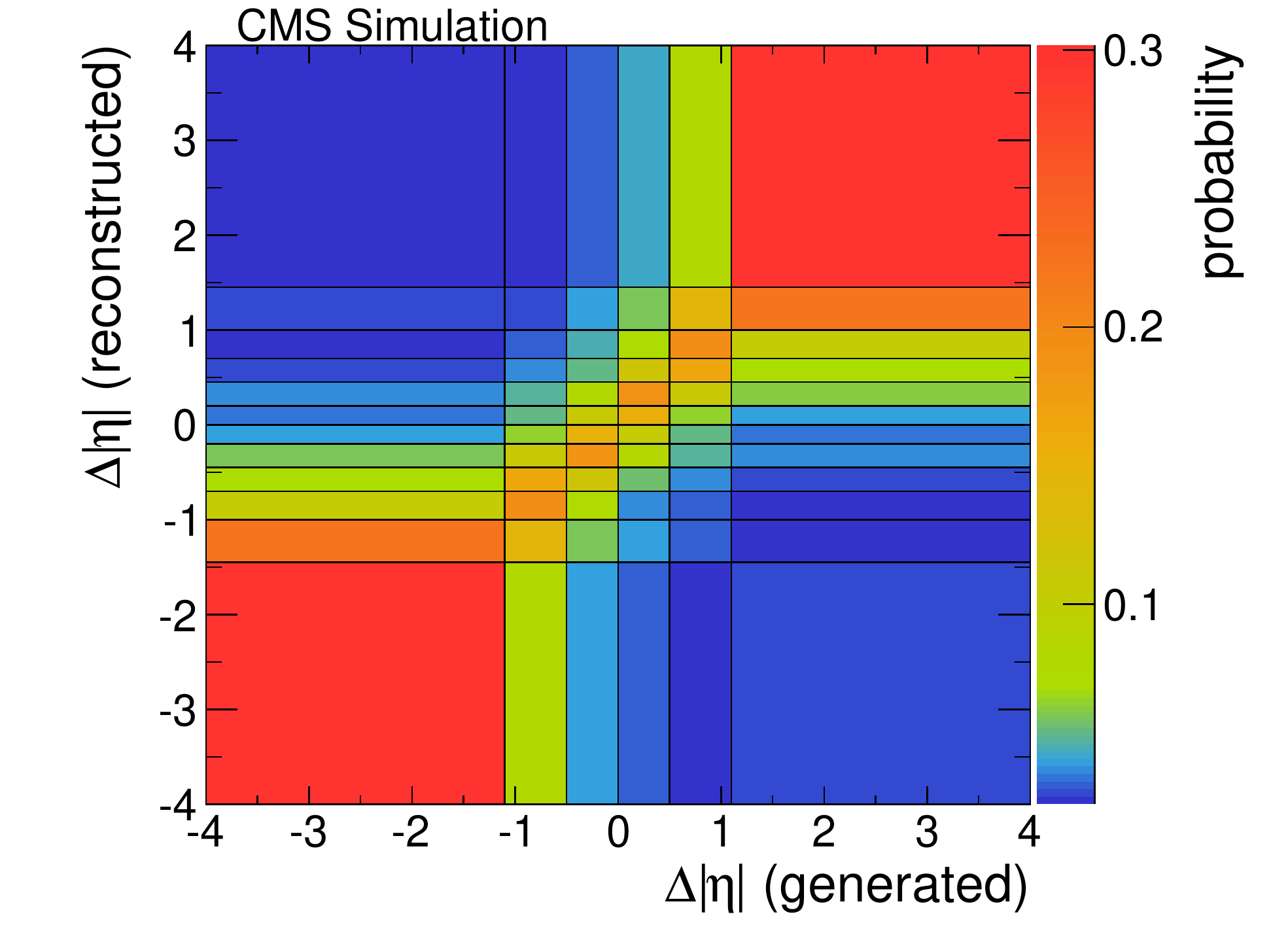}
         \caption{(left) Selection efficiency as a function of generated $\DeltaEta$, defined with respect to inclusive \ttbar production. (right) Migration matrix between the true (generated) and the reconstructed values in $\DeltaEta$, after the event selection.}
 \label{fig:Smearmatrix}
\end{figure*}

Distortions of the remaining \ttbar distributions relative to the true distributions can be factorized into effects from the event selection and event reconstruction. The values of $\DeltaEta$ or $\Delta y^{2}$ affect the probability for any event to survive all selection criteria (see Fig.~\ref{fig:Smearmatrix}~(left)), and thereby the distributions even before reconstruction. Further distortions can occur because of ambiguities in the assignment of jets to top-quark candidates, the determination of the neutrino momentum from the \PW-boson mass constraint, the energy resolution of the calorimeters and jet reconstruction, and the overall detector acceptance. The migration matrix, obtained from simulated \ttbar events and shown graphically in Fig.~\ref{fig:Smearmatrix}~(right), describes the migration of selected events from true values of $\DeltaEta$ to different reconstructed values.

To correct for the above effects, we apply a regularized unfolding procedure to the data~\cite{Blobel:2002pu} through a generalized matrix-inversion method.
The measured spectrum, denoted as vector $\vec{w}$, is divided into 12 bins (see $y$ axis of Fig.~\ref{fig:Smearmatrix}~(right)), where the bin widths are chosen to contain approximately an equal number of events. Six bins are used for the unfolded spectrum $\vec{x}$ (see $x$ axis of Fig.~\ref{fig:Smearmatrix}~(right)). Using twice as many bins for the uncorrected as for the corrected spectrum is recommended for this type of unfolding technique~\cite{Blobel:2002pu}.

The smearing matrix $S$, which accounts for migration and efficiency, is derived from simulated \ttbar events. Mathematically, this matrix is the product of the migration matrix, depicted in Fig.~\ref{fig:Smearmatrix}~(right), and a diagonal matrix with the efficiencies for each of the bins in Fig.~\ref{fig:Smearmatrix}~(left) on the diagonal, and all other elements set to zero. It defines the translation of the true spectrum $\vec{x}$ into the measured spectrum  $\vec{w} = S\vec{x}$. We solve this equation for the true spectrum $\vec{x}$ using a least-squares (LS) technique and searching for the $\vec{x}_{\mathrm{LS}}$ that minimizes the LS through the use of the generalized inverse of the smearing matrix $S$.

In general, the resulting solutions are unstable, with unacceptable fluctuations for small changes in $\vec{w}$. To regularize the problem and avoid unphysical fluctuations, two additional terms, a regularization term and a normalization term, are introduced in the procedure~\cite{Tikhonov,321114}. For both the $\DeltaEta$ and the $\Delta y^{2}$ variables, we use independent unfolding procedures based on the respective observable.

The performance of the unfolding algorithm is tested in sets of pseudoexperiments, each of which provides a randomly-generated sample distribution. The number of events from each contributing process is determined through a random number from a Gaussian distribution centred around the measured event rate given in Table~\ref{tab:FitResults}, with a width corresponding to the respective uncertainty. To take statistical variations into account, the number of expected events is defined by a Poisson distribution around the chosen Gaussian means. This final number of events for each process is drawn randomly from the appropriately simulated events to generate distributions for each pseudoexperiment. Each generated distribution is then subjected to the unfolding procedure described above. For all pseudoexperiments, we subtract the same number of background events as found in data.

We perform 50\,000 pseudoexperiments and compare the unfolded spectrum with the generated distribution in each experiment. The average asymmetry from these pseudoexperiments agrees well with the true asymmetry in the sample used to model the signal component and the pull distributions agree with expectations, indicating that the treatment of uncertainties is consistent with Gaussian behaviour. To test the unfolding procedure for different asymmetries, we reweight the events of the default $\mathrm{t \bar{t}}$ sample according to their $\DeltaEta$ or $\Delta y^{2}$ value, to artificially introduce asymmetries between $-0.2$ and $+0.2$, and then perform 50\,000 pseudoexperiments for each of the reweighted distributions. We find a linear dependence of the ensemble mean on the input value. While for $\DeltaEta$ the agreement is excellent, for $\Delta y^{2}$ we observe a slope for the linear dependence of 0.94 instead of 1.0, necessitating a correction of $1/0.94$ to the measured asymmetry. The statistical uncertainties of the measurements are found to be independent of the generated asymmetries.

\section{Estimation of systematic uncertainties}
\label{sec:Systematics}

The measured charge asymmetry $A_{C}$ can be affected by several sources of systematic uncertainty. Influences on the direction of the reconstructed top-quark momenta can change the value of the reconstructed charge asymmetry. Systematic uncertainties with an impact on the differential selection efficiency can also bias the result, while the overall selection efficiency and acceptance may not.
Variations in the background rates can also change the asymmetry attributed to the signal.
Since the uncorrected asymmetries observed in data are close to zero, such changes have only a small influence.
To evaluate each source of systematic uncertainty, we perform studies on pseudoexperiments using samples with systematically shifted parameters, and unfolding the distributions of interest as done with data.

The corrections on jet-energy scale (JES) and jet-energy resolution (JER) are changed by $\pm 1$ standard deviations of their $\eta$ and \pt-dependent uncertainties to estimate their effects on the measurement. Similarly, to estimate differences between simulation and data, other dedicated \ttbar samples are generated using different renormalization and factorization scales ($Q^{2}$), jet-matching thresholds~\cite{MLM}, and initial and final-state radiation (ISR and FSR). The systematic uncertainties on the measured asymmetry from the choice of parton distribution functions (PDF) for the colliding protons are estimated using the CTEQ6.6~\cite{cteq66} PDF set and the LHAPDF~\cite{lhapdf} package. In addition, the impact of the uncertainty on \cPqb-tagging efficiency, lepton selections, and lepton-trigger efficiencies are examined, taking their $\eta$ dependence into account. The uncertainty on the multijet background obtained from data, and on the frequency of occurrence of pile-up events, are also estimated. A mismodelling of \ETm and M3 in the simulation could change the estimation of the background rates. The measured charge asymmetry is found to be stable under such variations, as verified by shifting the amount of background in the pseudoexperiments. The estimations of all the systematic uncertainties are summarized in Table~\ref{tab:Syst}. The largest contribution arises from the jet-matching threshold, which is changed by factors of 2 and 0.5. Other important effects are from uncertainties in the $Q^2$ scale, from ISR and FSR in the simulated \ttbar sample, and from the uncertainty in the multijet-background model.

\newcommand\T{\rule{0pt}{2.6ex}}
\newcommand\B{\rule[-1.2ex]{0pt}{0pt}}
\begin{table*}
\caption{\label{tab:Syst}
 Listed are the positive and negative shifts on $A_C$ induced by systematic uncertainties in the pseudoexperiments from the different sources and the total.}
 \begin{center}
    \begin{tabular}{rrrrr} \hline
\T \B & \multicolumn{2}{c}{ $A_{C}^{\eta}$}      & \multicolumn{2}{c}{ $A_{C}^{y}$}\\  \hline
     Source                    & $-$ Shift & $+$ Shift & $-$ Shift   & $+$ Shift \\\hline
     JES                       & $-0.003$ & $ 0.000$    & $-0.007$    & $ 0.000$\\
     JER                       & $-0.002$ & $ 0.000$    & $-0.001$    & $+0.001$\\
     $Q^2$ scale               & $-0.005$ & $+0.008$    & $-0.013$    &$ 0.000$\\
     ISR/FSR                   & $-0.006$ & $+0.003$    & $ 0.000$    &$+0.024$\\
     Jet-matching threshold        & $-0.034$ & $+0.021$    & $-0.026$    &$+0.014$\\
     PDF                       & $-0.001$ & $+0.001$    & $-0.001$    &$+0.001$ \\
     \cPqb-tagging efficiency      & $-0.001$ & $+0.003$    & $ 0.000$    &$+0.001$\\
     Lepton ID/sel. efficiency & $-0.002$ & $+0.004$    & $-0.002$    &$+0.003$\\
     Multijet-background model & $-0.008$ & $+0.008$    & $-0.006$    &$+0.006$\\
     Pile-up                   & $-0.002$ & $+0.002$    & $ 0.000$    &$ 0.000$\\\hline
     Total                     & $-0.036$ & $+0.025$    & $-0.031$    &$+0.029$\\\hline
 \end{tabular}
\end{center}
\end{table*}

\section{Results}
\label{sec:Results}

\begin{figure*}[htbp]
 \centering
   \includegraphics[width=0.49\textwidth]{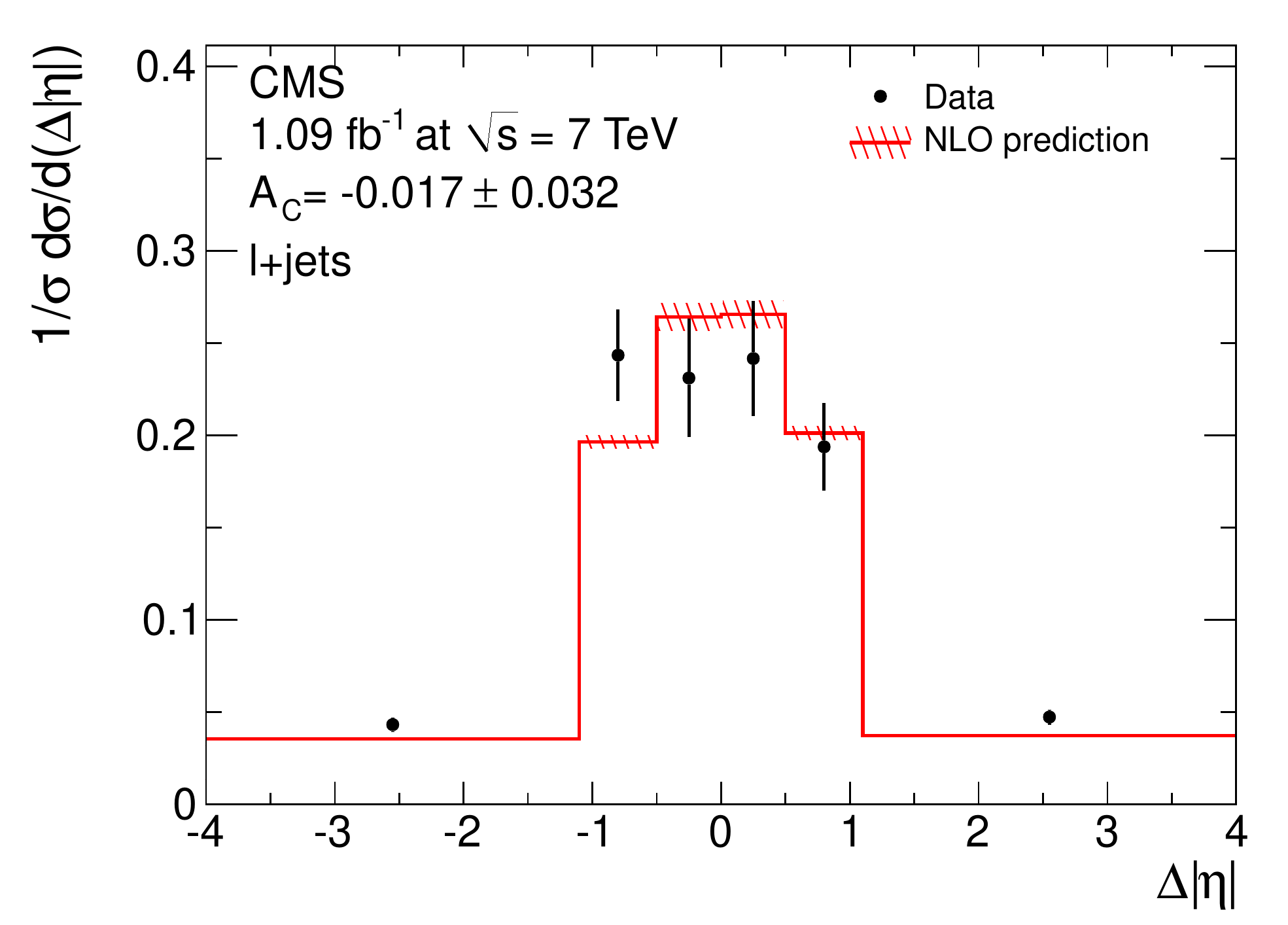}
   \includegraphics[width=0.49\textwidth]{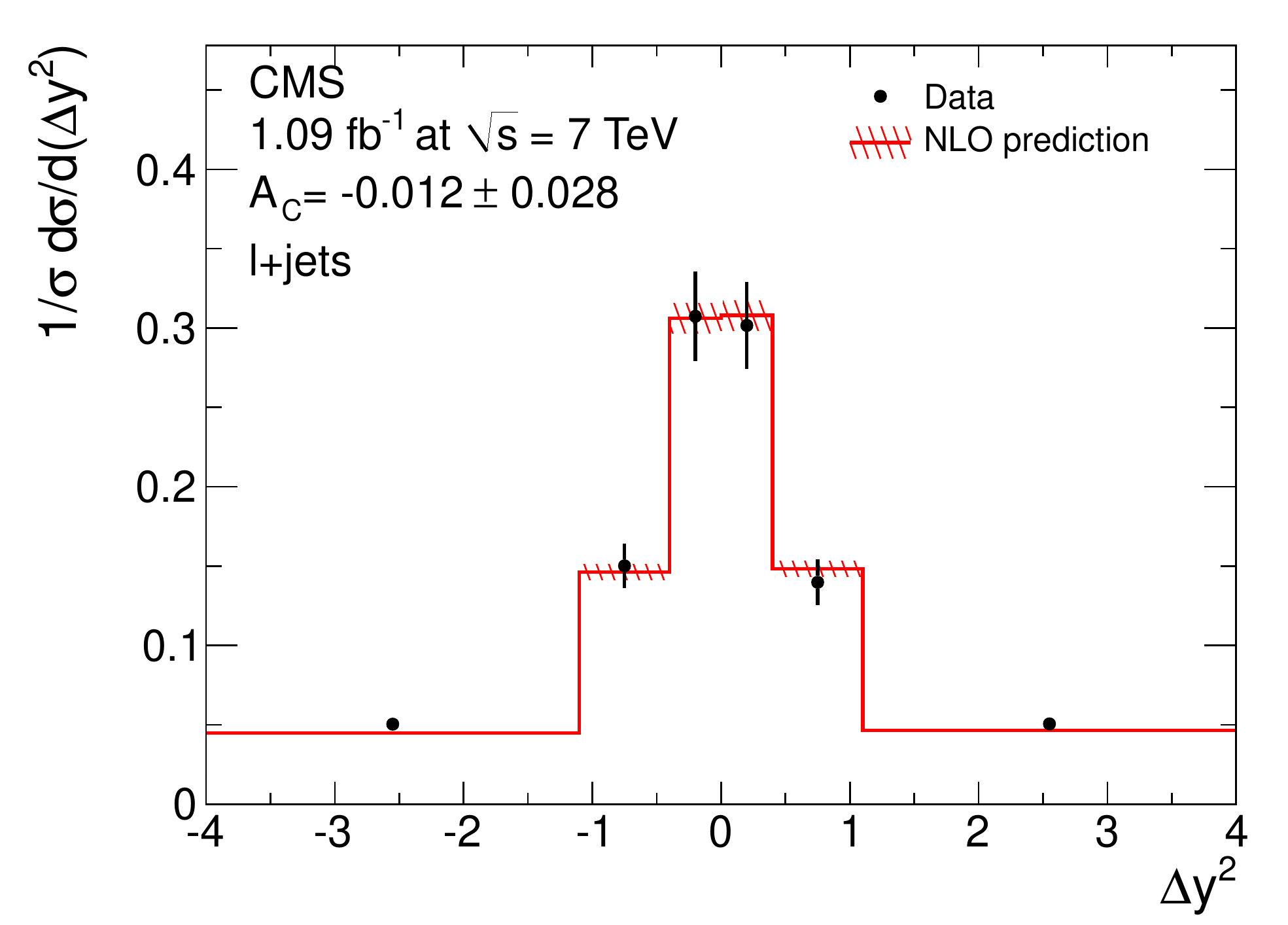}
       \caption{Unfolded $\DeltaEta$ (left) and $\Delta y^{2}$ (right) normalized spectra. The NLO prediction is based on the calculations of Ref.~\cite{Kuhn:2011ri}. The last bins include the sum of all contributions for $\left|\DeltaEta\right| > 4.0$ and $\left|\Delta y^{2}\right| > 4.0$, respectively. The uncertainties shown on the data are statistical, while the uncertainties on the prediction account also for the dependence on the top-quark mass, PDF, and factorization and renormalization scales.}
 \label{fig:Asy_Results}
\end{figure*}

Table~\ref{tab:Results} gives the values of the uncorrected asymmetries, the asymmetries after background subtraction (BG-subtracted), and the final, corrected asymmetries for both variables, along with the predicted theoretical values. Figure~\ref{fig:Asy_Results} shows the unfolded spectra used for computing the asymmetries, together with the SM prediction at NLO.

\renewcommand{\arraystretch}{1.2}
\begin{table*}[htbp]
\caption{\label{tab:Results} The measured asymmetries for both observables at the different stages of the analysis and the corresponding theory predictions. The final result for $A_{C}^{y}$ is corrected for a small bias observed in the dependence of the reconstructed value on the true value.}
 \begin{center}
    \begin{tabular}{lcc} \hline
Asymmetry                    &  $A_{C}^{\eta}$                 &  $A_{C}^{y}$ \\ \hline
Uncorrected                  &  $-0.004 \pm 0.009$\,(stat.) &  $-0.004 \pm 0.009$\,(stat.) \\
BG-subtracted        &  $-0.009 \pm 0.010$\,(stat.) &  $-0.007 \pm 0.010$\,(stat.) \\
Final corrected              &  $ -0.017 \pm 0.032~(\mathrm{stat.})\,^{+0.025}_{-0.036}~(\mathrm{syst.})$ & $-0.013 \pm 0.028~(\mathrm{stat.})\,^{+0.029}_{-0.031}~(\mathrm{syst.})$ \\ \hline
Theory predictions             &  $0.0136\pm 0.0008$          &   $0.0115\pm 0.0006$\\ \hline
 \end{tabular}
\end{center}
\end{table*}
\renewcommand{\arraystretch}{1.0}

\begin{figure*}[htbp]
 \centering
   \includegraphics[width=0.49\textwidth]{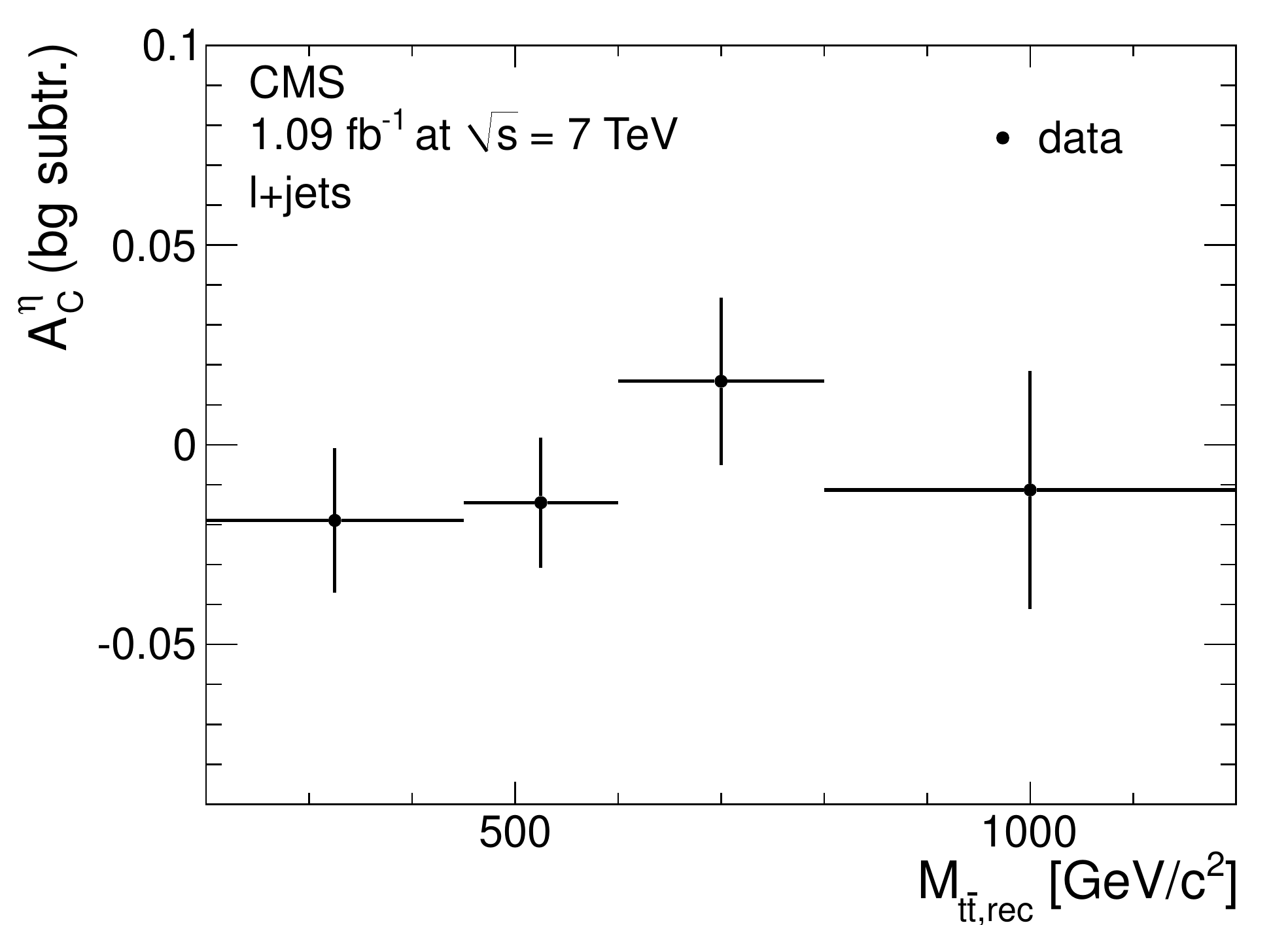}
   \includegraphics[width=0.49\textwidth]{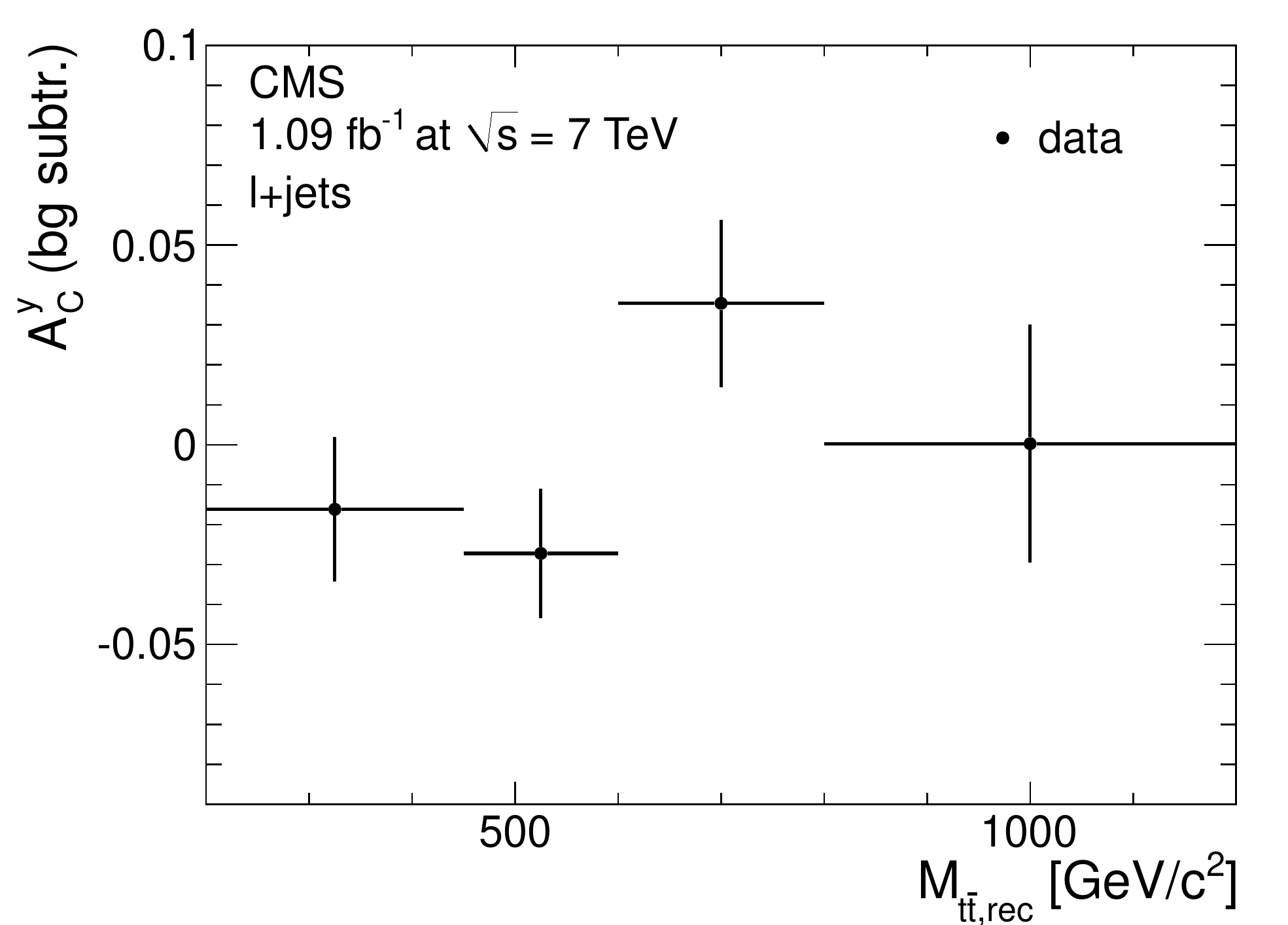}
   \caption{Background-subtracted asymmetries for $\DeltaEta$ (left) and $\Delta y^{2}$ (right) as functions of the reconstructed \ttbar invariant mass.}
 \label{fig:Mttbar}
\end{figure*}

Both measurements of the charge asymmetry are in agreement with the NLO predictions. We also measure the background-subtracted asymmetry as a function of the reconstructed \ttbar invariant mass. A dependence of the charge asymmetry on $M_{\ttbar}$, as large as reported by the CDF experiment~\cite{:2011kc}, could be visible in the reconstructed quantities, even without unfolding of $\DeltaEta$ (or $\Delta y^{2}$) and $M_{\ttbar}$.
Figure~\ref{fig:Mttbar} shows the results for the two variables, where no significant change in asymmetry is observed as a function of $M_{\ttbar}$ in the distributions before unfolding.

\section{Summary}

A measurement of the charge asymmetry in \ttbar production using data corresponding to an integrated luminosity of 1.09\fbinv has been reported. Events with top-quark pairs decaying in the $\text{lepton}{+}\text{jets}$ channel were selected and a full \ttbar event reconstruction was performed to determine the four-momenta of the top quarks and antiquarks. The measured distributions of the sensitive observables were then subjected to a regularized unfolding procedure to extract the asymmetry values, corrected for acceptance and reconstruction. The measured asymmetries are
\begin{equation}
A_{C}^{\eta} = -0.017 \pm 0.032~(\mathrm{stat.})\,^{+0.025}_{-0.036}~(\mathrm{syst.})~,
\end{equation}
and
\begin{equation}
A_{C}^{y} = -0.013 \pm 0.028~(\mathrm{stat.})\,^{+0.029}_{-0.031}~(\mathrm{syst.})~,
\end{equation}
consistent with SM predictions. The background-subtracted asymmetry shows no statistically significant dependence on the reconstructed \ttbar invariant mass.

\section*{Acknowledgment}
\hyphenation{Bundes-ministerium Forschungs-gemeinschaft Forschungs-zentren}We thank Johann H. K\"{u}hn and German Rodrigo for very fruitful discussions, and we wish to congratulate our colleagues in the CERN accelerator departments for the excellent performance of the LHC machine. We thank the technical and administrative staff at CERN and other CMS institutes, and acknowledge support from: FMSR (Austria); FNRS and FWO (Belgium); CNPq, CAPES, FAPERJ, and FAPESP (Brazil); MES (Bulgaria); CERN; CAS, MoST, and NSFC (China); COLCIENCIAS (Colombia); MSES (Croatia); RPF (Cyprus); Academy of Sciences and NICPB (Estonia); Academy of Finland, MEC, and HIP (Finland); CEA and CNRS/IN2P3 (France); BMBF, DFG, and HGF (Germany); GSRT (Greece); OTKA and NKTH (Hungary); DAE and DST (India); IPM (Iran); SFI (Ireland); INFN (Italy); NRF and WCU (Korea); LAS (Lithuania); CINVESTAV, CONACYT, SEP, and UASLP-FAI (Mexico); MSI (New Zealand); PAEC (Pakistan); SCSR (Poland); FCT (Portugal); JINR (Armenia, Belarus, Georgia, Ukraine, Uzbekistan); MST, MAE and RFBR (Russia); MSTD (Serbia); MICINN and CPAN (Spain); Swiss Funding Agencies (Switzerland); NSC (Taipei); TUBITAK and TAEK (Turkey); STFC (United Kingdom); DOE and NSF (USA).

Individuals have received support from the Marie-Curie programme and the European Research Council (European Union); the Leventis Foundation; the A. P. Sloan Foundation; the Alexander von Humboldt Foundation; the Belgian Federal Science Policy Office; the Fonds pour la Formation \`a la Recherche dans l'Industrie et dans l'Agriculture (FRIA-Belgium); the Agentschap voor Innovatie door Wetenschap en Technologie (IWT-Belgium); and the Council of Science and Industrial Research, India.

\bibliography{auto_generated}

\providecommand{\href}[2]{#2}\begingroup\raggedright\begin{thebibliography}{10}%
\makeatletter
\providecommand{\hrefCMSnoop }[0]{\@secondoftwo}%
\makeatother

\bibitem{Antunano:2007da}
\hrefCMSnoop {} {O.~Antunano, J.~H. K{\"u}hn, and G.~Rodrigo, ``Top quarks,
  axigluons and charge asymmetries at hadron colliders'',} \textit{ Phys. Rev.
  D} \textbf{ 77} (2008) 014003,
  \href{http://www.arXiv.org/abs/0709.1652}{\texttt{ arXiv:0709.1652}}.
\href{http://dx.doi.org/10.1103/PhysRevD.77.014003}{\texttt{
  doi:10.1103/PhysRevD.77.014003}}.
%%CITATION = 0709.1652;%%.

\bibitem{Frampton:2009rk}
\hrefCMSnoop {} {P.~H. Frampton, J.~Shu, and K.~Wang, ``Axigluon as possible
  explanation for ${\rm p\bar{p} \to t\bar{t}}$ forward-backward asymmetry'',}
  \textit{ Phys. Lett. B} \textbf{ 683} (2010) 294,
  \href{http://www.arXiv.org/abs/0911.2955}{\texttt{ arXiv:0911.2955}}.
\href{http://dx.doi.org/10.1016/j.physletb.2009.12.043}{\texttt{
  doi:10.1016/j.physletb.2009.12.043}}.
%%CITATION = 0911.2955;%%.

\bibitem{Rosner:1996eb}
\hrefCMSnoop {} {J.~L. Rosner, ``Prominent decay modes of a leptophobic
  {$Z^\prime$}'',} \textit{ Phys. Lett. B} \textbf{ 387} (1996) 113,
  \href{http://www.arXiv.org/abs/hep-ph/9607207}{\texttt{
  arXiv:hep-ph/9607207}}.
\href{http://dx.doi.org/10.1016/0370-2693(96)01022-2}{\texttt{
  doi:10.1016/0370-2693(96)01022-2}}.
%%CITATION = HEP-PH/9607207;%%.

\bibitem{Ferrario:2008wm}
\hrefCMSnoop {} {P.~Ferrario and G.~Rodrigo, ``Massive color-octet bosons and
  the charge asymmetries of top quarks at hadron colliders'',} \textit{ Phys.
  Rev. D} \textbf{ 78} (2008) 094018,
  \href{http://www.arXiv.org/abs/0809.3354}{\texttt{ arXiv:0809.3354}}.
\href{http://dx.doi.org/10.1103/PhysRevD.78.094018}{\texttt{
  doi:10.1103/PhysRevD.78.094018}}.
%%CITATION = 0809.3354;%%.

\bibitem{Ferrario:2009ee}
\hrefCMSnoop {} {P.~Ferrario and G.~Rodrigo, ``Heavy colored resonances in
  $t\overline{t} + \text{jet}$ at the {LHC}'',} \textit{ JHEP} \textbf{ 02}
  (2010) 051, \href{http://www.arXiv.org/abs/0912.0687}{\texttt{
  arXiv:0912.0687}}.
\href{http://dx.doi.org/10.1007/JHEP02(2010)051}{\texttt{
  doi:10.1007/JHEP02(2010)051}}.
%%CITATION = 0912.0687;%%.

\bibitem{AguilarSaavedra:2011hz}
\hrefCMSnoop {} {J.~A. Aguilar-Saavedra and M.~Perez-Victoria, ``Asymmetries in
  $t \bar{t}$ production: {LHC} versus {Tevatron}'',} (2011).
\href{http://www.arXiv.org/abs/1105.4606}{\texttt{ arXiv:1105.4606}}.
%%CITATION = 1105.4606;%%.

\bibitem{Kuhn:1998jr}
\hrefCMSnoop {} {J.~H. K{\"u}hn and G.~Rodrigo, ``Charge Asymmetry in
  Hadroproduction of Heavy Quarks'',} \textit{ Phys. Rev. Lett.} \textbf{ 81}
  (1998) 49, \href{http://www.arXiv.org/abs/hep-ph/9802268}{\texttt{
  arXiv:hep-ph/9802268}}.
\href{http://dx.doi.org/10.1103/PhysRevLett.81.49}{\texttt{
  doi:10.1103/PhysRevLett.81.49}}.
%%CITATION = HEP-PH/9802268;%%.

\bibitem{Kuhn:1998kw}
\hrefCMSnoop {} {J.~H. K{\"u}hn and G.~Rodrigo, ``Charge asymmetry of heavy
  quarks at hadron colliders'',} \textit{ Phys. Rev. D} \textbf{ 59} (1999)
  054017, \href{http://www.arXiv.org/abs/hep-ph/9807420}{\texttt{
  arXiv:hep-ph/9807420}}.
\href{http://dx.doi.org/10.1103/PhysRevD.59.054017}{\texttt{
  doi:10.1103/PhysRevD.59.054017}}.
%%CITATION = HEP-PH/9807420;%%.

\bibitem{:2011kc}
\hrefCMSnoop {} {{ CDF} Collaboration, ``Evidence for a mass dependent
  forward-backward asymmetry in top quark pair production'',} \textit{ Phys.
  Rev. D} \textbf{ 83} (2011) 112003,
  \href{http://www.arXiv.org/abs/1101.0034}{\texttt{ arXiv:1101.0034}}.
\href{http://dx.doi.org/10.1103/PhysRevD.83.112003}{\texttt{
  doi:10.1103/PhysRevD.83.112003}}.
%%CITATION = 1101.0034;%%.

\bibitem{PhysRevD.84.112005}
\hrefCMSnoop {} {{ D0} Collaboration, ``Forward-backward asymmetry in top
  quark-antiquark production'',} \textit{ Phys. Rev. D} \textbf{ 84} (2011)
  112005. \href{http://dx.doi.org/10.1103/PhysRevD.84.112005}{\texttt{
  doi:10.1103/PhysRevD.84.112005}}.

\bibitem{Bernreuther:2010ny}
\hrefCMSnoop {} {W.~Bernreuther and Z.-G. Si, ``Distributions and correlations
  for top quark pair production and decay at the {T}evatron and {LHC}'',}
  \textit{ Nucl. Phys. B} \textbf{ 837} (2010) 90,
  \href{http://www.arXiv.org/abs/1003.3926}{\texttt{ arXiv:1003.3926}}.
\href{http://dx.doi.org/10.1016/j.nuclphysb.2010.05.001}{\texttt{
  doi:10.1016/j.nuclphysb.2010.05.001}}.
%%CITATION = 1003.3926;%%.

\bibitem{Kuhn:2011ri}
\hrefCMSnoop {} {J.~H. K{\"u}hn and G.~Rodrigo, ``Charge asymmetries of top
  quarks at hadron colliders revisited'',} (2011).
\href{http://www.arXiv.org/abs/1109.6830}{\texttt{ arXiv:1109.6830}}.
%%CITATION = 1109.6830;%%.

\bibitem{Gabrielli:2011jf}
\hrefCMSnoop {} {E.~Gabrielli and M.~Raidal, ``Effective axial-vector coupling
  of gluon as an explanation to the top quark asymmetry'',} \textit{ Phys. Rev.
  D} \textbf{ 84} (2011) 054017,
  \href{http://www.arXiv.org/abs/1106.4553}{\texttt{ arXiv:1106.4553}}.
  \href{http://dx.doi.org/10.1103/PhysRevD.84.054017}{\texttt{
  doi:10.1103/PhysRevD.84.054017}}.

\bibitem{Bauer:2010iq}
\hrefCMSnoop {} {M.~Bauer, F.~Goertz, U.~Haisch{ et~al.}, ``Top-quark
  forward-backward asymmetry in {Randall-Sundrum} models beyond the leading
  order'',} \textit{ JHEP} \textbf{ 11} (2010) 039,
  \href{http://www.arXiv.org/abs/1008.0742}{\texttt{ arXiv:1008.0742}}.
\href{http://dx.doi.org/10.1007/JHEP11(2010)039}{\texttt{
  doi:10.1007/JHEP11(2010)039}}.
%%CITATION = 1008.0742;%%.

\bibitem{Cao:2009uz}
\hrefCMSnoop {} {J.~Cao, Z.~Heng, L.~Wu{ et~al.}, ``Top quark forward-backward
  asymmetry at the {Tevatron}: A comparative study in different new physics
  models'',} \textit{ Phys. Rev. D} \textbf{ 81} (2010) 014016,
  \href{http://www.arXiv.org/abs/0912.1447}{\texttt{ arXiv:0912.1447}}.
\href{http://dx.doi.org/10.1103/PhysRevD.81.014016}{\texttt{
  doi:10.1103/PhysRevD.81.014016}}.
%%CITATION = 0912.1447;%%.

\bibitem{Shu:2009xf}
\hrefCMSnoop {} {J.~Shu, T.~M.~P. Tait, and K.~Wang, ``Explorations of the top
  quark forward-backward asymmetry at the {Tevatron}'',} \textit{ Phys. Rev. D}
  \textbf{ 81} (2010) 034012,
  \href{http://www.arXiv.org/abs/0911.3237}{\texttt{ arXiv:0911.3237}}.
\href{http://dx.doi.org/10.1103/PhysRevD.81.034012}{\texttt{
  doi:10.1103/PhysRevD.81.034012}}.
%%CITATION = 0911.3237;%%.

\bibitem{Arhrib:2009hu}
\hrefCMSnoop {} {A.~Arhrib, R.~Benbrik, and C.-H. Chen, ``Forward-backward
  asymmetry of top quark in diquark models'',} \textit{ Phys. Rev. D} \textbf{
  82} (2010) 034034, \href{http://www.arXiv.org/abs/0911.4875}{\texttt{
  arXiv:0911.4875}}.
\href{http://dx.doi.org/10.1103/PhysRevD.82.034034}{\texttt{
  doi:10.1103/PhysRevD.82.034034}}.
%%CITATION = 0911.4875;%%.

\bibitem{Dorsner:2009mq}
\hrefCMSnoop {} {I.~Dor{\v{s}}ner, S.~Fajfer, J.~F. Kamenik{ et~al.}, ``Light
  colored scalars from grand unification and the forward-backward asymmetry in
  $t\overline{t}$ quark pair production'',} \textit{ Phys. Rev. D} \textbf{ 81}
  (2010) 055009, \href{http://www.arXiv.org/abs/0912.0972}{\texttt{
  arXiv:0912.0972}}.
\href{http://dx.doi.org/10.1103/PhysRevD.81.055009}{\texttt{
  doi:10.1103/PhysRevD.81.055009}}.
%%CITATION = 0912.0972;%%.

\bibitem{Djouadi:2009nb}
\hrefCMSnoop {} {A.~Djouadi, G.~Moreau, F.~Richard{ et~al.}, ``The
  forward-backward asymmetry of top quark production at the {Tevatron} in
  warped extra dimensional models'',} \textit{ Phys. Rev. D} \textbf{ 82}
  (2010) 071702, \href{http://www.arXiv.org/abs/0906.0604}{\texttt{
  arXiv:0906.0604}}.
\href{http://dx.doi.org/10.1103/PhysRevD.82.071702}{\texttt{
  doi:10.1103/PhysRevD.82.071702}}.
%%CITATION = 0906.0604;%%.

\bibitem{Djouadi:2011aj}
\hrefCMSnoop {} {A.~Djouadi, G.~Moreau, and F.~Richard, ``Forward-backward
  asymmetries of the bottom and top quarks in warped extra-dimensional models:
  {LHC} predictions from the {LEP} and {T}evatron anomalies'',} \textit{ Phys.
  Lett. B} \textbf{ 701} (2011) 458,
  \href{http://www.arXiv.org/abs/1105.3158}{\texttt{ arXiv:1105.3158}}.
\href{http://dx.doi.org/10.1016/j.physletb.2011.06.028}{\texttt{
  doi:10.1016/j.physletb.2011.06.028}}.
%%CITATION = 1105.3158;%%.

\bibitem{Rodrigo:2010gm}
\hrefCMSnoop {} {G.~Rodrigo and P.~Ferrario, ``Charge asymmetry: A theory
  appraisal'',} \textit{ Nuovo Cim. C} \textbf{ 33} (2010) 221,
  \href{http://www.arXiv.org/abs/1007.4328}{\texttt{ arXiv:1007.4328}}.
  \href{http://dx.doi.org/10.1393/ncc/i2010-10646-5}{\texttt{
  doi:10.1393/ncc/i2010-10646-5}}.

\bibitem{Martynov:2010ed}
\hrefCMSnoop {} {M.~V. Martynov and A.~D. Smirnov, ``On mass limit for chiral
  color symmetry {G$'$}-boson from {Tevatron} data on $t \bar{t}$
  production'',} \textit{ Mod. Phys. Lett. A} \textbf{ 25} (2010) 2637,
  \href{http://www.arXiv.org/abs/1006.4246}{\texttt{ arXiv:1006.4246}}.
\href{http://dx.doi.org/10.1142/S0217732310033931}{\texttt{
  doi:10.1142/S0217732310033931}}.
%%CITATION = 1006.4246;%%.

\bibitem{AguilarSaavedra:2011vw}
\hrefCMSnoop {} {J.~A. Aguilar-Saavedra and M.~Perez-Victoria, ``Probing the
  {Tevatron} $t \bar{t}$ asymmetry at {LHC}'',} \textit{ JHEP} \textbf{ 05}
  (2011) 034, \href{http://www.arXiv.org/abs/1103.2765}{\texttt{
  arXiv:1103.2765}}.
\href{http://dx.doi.org/10.1007/JHEP05(2011)034}{\texttt{
  doi:10.1007/JHEP05(2011)034}}.
%%CITATION = 1103.2765;%%.

\bibitem{Jung:2009pi}
\hrefCMSnoop {} {D.-W. Jung, P.~Ko, J.~S. Lee{ et~al.}, ``Model independent
  analysis of the forward-backward asymmetry of top quark production at the
  {Tevatron}'',} \textit{ Phys. Lett. B} \textbf{ 691} (2010) 238,
  \href{http://www.arXiv.org/abs/0912.1105}{\texttt{ arXiv:0912.1105}}.
\href{http://dx.doi.org/10.1016/j.physletb.2010.06.040}{\texttt{
  doi:10.1016/j.physletb.2010.06.040}}.
%%CITATION = 0912.1105;%%.

\bibitem{Cao:2010zb}
\hrefCMSnoop {} {Q.-H. Cao, D.~McKeen, J.~L. Rosner{ et~al.},
  ``Forward-backward asymmetry of top quark pair production'',} \textit{ Phys.
  Rev. D} \textbf{ 81} (2010) 114004,
  \href{http://www.arXiv.org/abs/1003.3461}{\texttt{ arXiv:1003.3461}}.
\href{http://dx.doi.org/10.1103/PhysRevD.81.114004}{\texttt{
  doi:10.1103/PhysRevD.81.114004}}.
%%CITATION = 1003.3461;%%.

\bibitem{Bai:2011ed}
\hrefCMSnoop {} {Y.~Bai, J.~L. Hewett, J.~Kaplan{ et~al.}, ``{LHC} predictions
  from a {T}evatron anomaly in the top quark forward-backward asymmetry'',}
  \textit{ JHEP} \textbf{ 03} (2011) 003,
  \href{http://www.arXiv.org/abs/1101.5203}{\texttt{ arXiv:1101.5203}}.
\href{http://dx.doi.org/10.1007/JHEP03(2011)003}{\texttt{
  doi:10.1007/JHEP03(2011)003}}.
%%CITATION = 1101.5203;%%.

\bibitem{Ligeti:2011vt}
\hrefCMSnoop {} {Z.~Ligeti, G.~M. Tavares, and M.~Schmaltz, ``Explaining the $t
  \bar{t}$ forward-backward asymmetry without dijet or flavor anomalies'',}
  \textit{ JHEP} \textbf{ 06} (2011) 109,
  \href{http://www.arXiv.org/abs/1103.2757}{\texttt{ arXiv:1103.2757}}.
\href{http://dx.doi.org/10.1007/JHEP06(2011)109}{\texttt{
  doi:10.1007/JHEP06(2011)109}}.
%%CITATION = 1103.2757;%%.

\bibitem{Diener:2009ee}
\hrefCMSnoop {} {R.~Diener, S.~Godfrey, and T.~A. Martin, ``Using final state
  pseudorapidities to improve s-channel resonance observables at the {LHC}'',}
  \textit{ Phys.Rev. D} \textbf{ 80} (2009) 075014,
  \href{http://www.arXiv.org/abs/0909.2022}{\texttt{ arXiv:0909.2022}}.
  \href{http://dx.doi.org/10.1103/PhysRevD.80.075014}{\texttt{
  doi:10.1103/PhysRevD.80.075014}}.

\bibitem{Jung:2011zv}
\hrefCMSnoop {} {S.~Jung, A.~Pierce, and J.~D. Wells, ``Top quark asymmetry
  from a non-{A}belian horizontal symmetry'',} \textit{ Phys. Rev. D} \textbf{
  83} (2011) 114039, \href{http://www.arXiv.org/abs/1103.4835}{\texttt{
  arXiv:1103.4835}}.
\href{http://dx.doi.org/10.1103/PhysRevD.83.114039}{\texttt{
  doi:10.1103/PhysRevD.83.114039}}.
%%CITATION = 1103.4835;%%.

\bibitem{cms}
\hrefCMSnoop {} {{ CMS} Collaboration, ``The {CMS} experiment at the {CERN}
  {LHC}'',} \textit{ JINST} \textbf{ 3} (2008) S08004.
\href{http://dx.doi.org/10.1088/1748-0221/3/08/S08004}{\texttt{
  doi:10.1088/1748-0221/3/08/S08004}}.
%%CITATION = JINST,3,S08004;%%.

\bibitem{madgraph}
\hrefCMSnoop {} {J.~Alwall, M.~Herquet, F.~Maltoni{ et~al.}, ``{MadGraph} 5:
  going beyond'',} \textit{ JHEP} \textbf{ 06} (2011) 128,
  \href{http://www.arXiv.org/abs/1106.0522}{\texttt{ arXiv:1106.0522}}.
  \href{http://dx.doi.org/10.1007/JHEP06(2011)128}{\texttt{
  doi:10.1007/JHEP06(2011)128}}.

\bibitem{pythia}
\hrefCMSnoop {} {T.~Sj\"{o}strand, S.~Mrenna, and P.~Skands, ``{PYTHIA} 6.4
  physics and manual'',}
  \href{http://www.arXiv.org/abs/hep-ph/0603175}{\texttt{
  arXiv:hep-ph/0603175}}.
  \href{http://dx.doi.org/10.1088/1126-6708/2006/05/026}{\texttt{
  doi:10.1088/1126-6708/2006/05/026}}.

\bibitem{MLM}
\hrefCMSnoop {} {M.~L. Mangano, M.~Moretti, F.~Piccinini{ et~al.}, ``{Matching
  matrix elements and shower evolution for top- quark production in hadronic
  collisions}'',} \textit{ JHEP} \textbf{ 01} (2007) 013,
  \href{http://www.arXiv.org/abs/hep-ph/0611129}{\texttt{
  arXiv:hep-ph/0611129}}.
\href{http://dx.doi.org/10.1088/1126-6708/2007/01/013}{\texttt{
  doi:10.1088/1126-6708/2007/01/013}}.
%%CITATION = HEP-PH/0611129;%%.

\bibitem{PFT-09-001}
\href {http://cdsweb.cern.ch/record/1194487} {{ CMS} Collaboration,
  ``Particle--Flow Event Reconstruction in {CMS} and Performance for Jets,
  Taus, and {\MET}'',} CMS Physics Analysis Summary CMS-PAS-PFT-09-001, (2009).

\bibitem{ElectronPAS}
\href {http://cdsweb.cern.ch/record/1299116} {{ CMS} Collaboration, ``Electron
  Reconstruction and Identification at $\sqrt{s} = 7\,${TeV}'',} CMS Physics
  Analysis Summary CMS-PAS-EGM-10-004, (2010).

\bibitem{antikt}
\hrefCMSnoop {} {M.~Cacciari, G.~P. Salam, and G.~Soyez, ``{The anti-$k_t$ jet
  clustering algorithm}'',} \textit{ JHEP} \textbf{ 04} (2008) 063,
  \href{http://www.arXiv.org/abs/0802.1189}{\texttt{ arXiv:0802.1189}}.
\href{http://dx.doi.org/10.1088/1126-6708/2008/04/063}{\texttt{
  doi:10.1088/1126-6708/2008/04/063}}.
%%CITATION = 0802.1189;%%.

\bibitem{fastjet}
\hrefCMSnoop {} {M.~Cacciari and G.~P. Salam, ``Dispelling the {$N^3$} myth for
  the $k_t$ jet-finder'',} \textit{ Phys. Lett. B} \textbf{ 641} (2006) 57,
  \href{http://www.arXiv.org/abs/hep-ph/0512210}{\texttt{
  arXiv:hep-ph/0512210}}.
\href{http://dx.doi.org/10.1016/j.physletb.2006.08.037}{\texttt{
  doi:10.1016/j.physletb.2006.08.037}}.
%%CITATION = HEP-PH/0512210;%%.

\bibitem{fastjet2}
\hrefCMSnoop {} {M.~Cacciari, G.~P. Salam, and G.~Soyez, ``{FastJet} user
  manual'',} (2011). \href{http://www.arXiv.org/abs/1111.6097}{\texttt{
  arXiv:1111.6097}}.

\bibitem{TCHE}
\href {http://cdsweb.cern.ch/record/1279144} {{ CMS} Collaboration,
  ``Commissioning of b-jet identification with pp collisions at $\sqrt{s}$ = 7
  {TeV}'',} CMS Physics Analysis Summary CMS-PAS-BTV-10-001, (2010).

\bibitem{BTAG}
\href {http://cdsweb.cern.ch/record/1366061} {{ CMS} Collaboration,
  ``Performance of b-jet identification in {CMS}'',} CMS Physics Analysis
  Summary CMS-PAS-BTV-11-001, (2011).

\bibitem{Chatrchyan:2011ew}
\hrefCMSnoop {} {{ CMS} Collaboration, ``Measurement of the top-antitop
  production cross section in pp collisions at $\sqrt s = 7\,$TeV using the
  kinematic properties of events with leptons and jets'',} \textit{ Eur. Phys.
  J. C} \textbf{ 71} (2011) 1721.
\href{http://dx.doi.org/10.1140/epjc/s10052-011-1721-3}{\texttt{
  doi:10.1140/epjc/s10052-011-1721-3}}.
%%CITATION = 1106.0902;%%.

\bibitem{Blobel:2002pu}
\hrefCMSnoop {} {V.~Blobel, ``An unfolding method for high energy physics
  experiments'',} (2002).
\href{http://www.arXiv.org/abs/hep-ex/0208022}{\texttt{ arXiv:hep-ex/0208022}}.
%%CITATION = HEP-EX/0208022;%%.

\bibitem{Tikhonov}
\hrefCMSnoop {} {A.~Tikhonov, ``Solution of incorrectly formulated problems and
  the regularization method'',} \textit{ Soviet Mathematics Doklady} \textbf{
  4} (1963) 1035.

\bibitem{321114}
\hrefCMSnoop {} {D.~L. Phillips, ``A Technique for the Numerical Solution of
  Certain Integral Equations of the First Kind'',} \textit{ J. ACM} \textbf{ 9}
  (1962) 84. \href{http://dx.doi.org/10.1145/321105.321114}{\texttt{
  doi:10.1145/321105.321114}}.

\bibitem{cteq66}
\hrefCMSnoop {} {P.~M. Nadolsky, H.-L. Lai, Q.-H. Cao{ et~al.}, ``Implications
  of {CTEQ} global analysis for collider observables'',} \textit{ Phys. Rev. D}
  \textbf{ 78} (2008) 013004,
  \href{http://www.arXiv.org/abs/0802.0007}{\texttt{ arXiv:0802.0007}}.
\href{http://dx.doi.org/10.1103/PhysRevD.78.013004}{\texttt{
  doi:10.1103/PhysRevD.78.013004}}.
%%CITATION = 0802.0007;%%.

\bibitem{lhapdf}
\hrefCMSnoop {} {{ WBG} Collaboration, ``The {L}es {H}ouches {A}ccord {PDFs}
  ({LHAPDF}) and {Lhaglue}'',} (2005).
  \href{http://www.arXiv.org/abs/hep-ph/0508110}{\texttt{
  arXiv:hep-ph/0508110}}.

\end{thebibliography}\endgroup

\cleardoublepage \appendix\section{The CMS Collaboration \label{app:collab}}\begin{sloppypar}\hyphenpenalty=5000\widowpenalty=500\clubpenalty=5000\textbf{Yerevan Physics Institute,  Yerevan,  Armenia}\\*[0pt]
S.~Chatrchyan, V.~Khachatryan, A.M.~Sirunyan, A.~Tumasyan
\vskip\cmsinstskip
\textbf{Institut f\"{u}r Hochenergiephysik der OeAW,  Wien,  Austria}\\*[0pt]
W.~Adam, T.~Bergauer, M.~Dragicevic, J.~Er\"{o}, C.~Fabjan, M.~Friedl, R.~Fr\"{u}hwirth, V.M.~Ghete, J.~Hammer\cmsAuthorMark{1}, M.~Hoch, N.~H\"{o}rmann, J.~Hrubec, M.~Jeitler, W.~Kiesenhofer, A.~Knapitsch, M.~Krammer, D.~Liko, I.~Mikulec, M.~Pernicka$^{\textrm{\dag}}$, B.~Rahbaran, C.~Rohringer, H.~Rohringer, R.~Sch\"{o}fbeck, J.~Strauss, A.~Taurok, F.~Teischinger, P.~Wagner, W.~Waltenberger, G.~Walzel, E.~Widl, C.-E.~Wulz
% ...
\vskip\cmsinstskip
\dag:~Deceased\\
1:~~Also at CERN, European Organization for Nuclear Research, Geneva, Switzerland\\
% ...

\end{sloppypar}
\end{document}